\documentclass{agujournal2019}
\usepackage{url} 
\usepackage{pdfpages}
\usepackage[inline]{trackchanges}
\usepackage{subfigure}
\usepackage{amsfonts, amsmath, amssymb}
\usepackage{algorithm, algpseudocode}
\usepackage{times}
\usepackage{multirow}
\usepackage{booktabs}
\usepackage[round]{natbib}
\usepackage[colorlinks=true, linkcolor=blue, citecolor=blue, urlcolor=blue]{hyperref}
\usepackage{graphicx}
\DeclareGraphicsExtensions{.pdf,.png,.jpg}
\usepackage{adjustbox}

\draftfalse

\journalname{JGR: Machine Learning and Computation}

\begin{document}

%
%


\title{An intercomparison of generative machine learning methods for downscaling precipitation at fine spatial scales}
%
%




\authors{Neelesh Rampal\affil{1}, Bryn Ward-Leikis\affil{2},  Yun Sing Koh\affil{2}, Peter B. Gibson\affil{1}, Hong-Yang Liu\affil{1,2}, Vassili Kitsios\affil{3,4}, Tristan Meyers\affil{1}, Jeff Adie\affil{5}, Yang Juntao\affil{5} \& Steven C. Sherwood\affil{6}}

\affiliation{1}{Earth Sciences New Zealand, New Zealand.}
\affiliation{2}{School of Computer Science, University of Auckland,  New Zealand.}
\affiliation{3}{Commonwealth Scientific and Industrial Research Organisation (CSIRO), Environment, 107-121 Station Street, Aspendale, 3195, Victoria, Australia.}
\affiliation{4}{Laboratory for Turbulence Research in Aerospace and Combustion, Department of Mechanical and Aerospace Engineering, Monash University, Clayton, 3800, Victoria, Australia}
\affiliation{6}{NVIDIA AI Technology Centre, NVIDIA Corporation, Singapore.}
\affiliation{6}{ARC Centre of Excellence for Weather of the 21st Century \& Climate Change Research Centre, University of New South Wales, Sydney, Australia}







\correspondingauthor{Neelesh Rampal}{neelesh.rampal@niwa.co.nz}




\begin{keypoints}
\item We benchmark GANs, diffusion, and flow matching for precipitation downscaling, evaluating structure, extremes, and climate change signals.
\item Flow matching and diffusion models rival GANs in skill, but yield better-calibrated ensembles and more realistic precipitation fields.
\item Most models underestimate the climate change signal of extremes despite good skill on other metrics, except one GAN and flow-matching model.
\end{keypoints}


%
%

%
%

\begin{abstract}
Machine learning (ML) offers a computationally efficient approach for generating large ensembles of high-resolution climate projections, but deterministic ML methods often smooth fine-scale structures and underestimate extremes. While stochastic generative models show promise, few studies have compared their skill under both present-day and future climates. This study compares Generative Adversarial Networks (GANs), flow matching and diffusion models across multiple configurations for downscaling daily precipitation from a regional climate model (RCM) over New Zealand. Model skill is assessed across spatial structure, distributional metrics, climatological means, extremes, ensemble calibration, and climate change signals. Unlike GANs, diffusion and flow matching models generate predictions through many sequential steps. Here we show that using higher-order differential equation solvers, the number of steps required can be reduced with only a minor reduction in skill, heavily reducing the computational burden for downscaling large ensembles, which may have otherwise prevented their use in operational settings. Overall, GANs, flow matching and diffusion perform competitively across most metrics, except that diffusion and flow matching produce higher-fidelity predictions and better-calibrated ensembles compared to GANs --which are under-dispersive. Most approaches capture mean precipitation signals reasonably well, but underestimate end-of-century climate change signals of extreme precipitation, despite being trained on RCM simulations spanning the future period. Only one GAN and one flow matching configuration can reproduce this change signal reliably. These results highlight the importance of evaluating model performance across a comprehensive set of metrics, and that neither visual realism nor good skill on standard metrics guarantee skill in predicting climate change signals.
\end{abstract}

\section*{Plain Language Summary}
Running high-resolution, physics-based climate simulations is computationally expensive. Machine learning (ML) offers a cheaper alternative for fine-scale climate simulations, but standard regression-based ML methods tend to produce overly smooth predictions that underestimate extremes. We tested three families of generative ML methods — generative adversarial networks (GANs), diffusion models, and flow matching models — for predicting fine-scale daily precipitation over New Zealand. We evaluated how well each method reproduced day-to-day weather, spatial structure, and extremes against high-resolution, physics-based simulations. Importantly, we assessed whether ML approaches can reliably project future changes in extreme precipitation — a critical test of their fitness-for-purpose for producing regional climate projections. Diffusion and flow matching models generate predictions through many sequential steps, making them slower than GANs; we show that improved mathematical solvers can drastically reduce this cost with little loss in skill. While diffusion and flow matching models are generally as skillful as GANs, they produce more realistic precipitation fields and more dispersive ensembles. However, most methods underestimated end-of-century changes in extreme precipitation, with only one GAN and one flow matching configuration capturing this reliably. Overall, a model that performs well on standard metrics does not necessarily produce reliable regional climate projections, underscoring the need for careful ML evaluation.

\section{Introduction}\label{section:introduction}







The typical spatial resolution of Global Climate Models (GCMs) ($\sim$ 100-150 km) is too coarse to anticipate future climate changes at local scales, where the impacts of climate change are experienced \citep{fowler2007linking, maraun2016bias}. To simulate future changes at local scales, Regional Climate Models (RCMs) downscale coarse-resolution climate projections to finer spatial scales, typically achieving resolutions of approximately 10–25 km in Coordinated Regional Climate Downscaling Experiment (CORDEX)-type experiments \citep{feser2011regional, rummukainen2010state}. However, running RCMs is computationally expensive, limiting the number of GCMs that can be downscaled \citep{ban2014evaluation, coppola2020first}. This has motivated the development of RCM emulators, which are significantly more computationally efficient than RCMs, and can be used to downscale large ensembles of GCMs—a capability necessary to better sample uncertainty (model, scenario, and internal variability) in climate projections \citep{rampal2024enhancing, rampal2025downscaling, lewis2025generative, lehner_origin_2023, deser_projecting_2014}. RCM emulators are empirical algorithms, typically based on statistical or machine learning methods, that learn the relationship between coarse-resolution GCM boundary conditions (predictor variables) and high-resolution climate fields (target variables) from existing RCM simulations \citep{chadwick2011artificial, holden2015emulation, rampal2022high, boe2023simple, doury2023regional, van_der_meer_deep_2023, rampal2024enhancing, maraun_value_2015}.

A wide variety of algorithms have been used for the emulation of RCM, ranging from traditional machine learning approaches \citep[e.g.,][]{holden2015emulation, chadwick2011artificial}, convolutional neural networks (CNNs) \citep[e.g.,][]{doury2023regional} and, more recently, generative AI algorithms \citep[e.g.,][]{addison2022machine}. A key limitation of regression-based deep learning methods is that training with Mean Squared Error (MSE) or similar loss functions causes models to ''regress-to-the-mean'', producing overly smooth outputs that underestimate fine-scale variability and extremes \citep{lopez2023global, izumi2022super, rampal2022high, rampal2025reliable, addison2022machine}. To overcome this issue, generative AI-based stochastic methods, such as generative adversarial networks (GANs), diffusion models and flow matching algorithms, have been used more recently for RCM emulation \citep{leinonen2021stochastic, price2022increasing, harris2022generative, wang2021fast, wetherell2026flow}. In particular, a key advantage of such methods is their ability to generate an arbitrary number of output samples from a single conditional input. This allows the creation of large ensembles of predictions, which can be used to estimate the model’s variance, dispersion, and overall uncertainty \citep{leinonen2021stochastic, bihlo2021generative, harris2022generative, schillinger2025enscale}.

GANs are a type of generative model \citep{goodfellow2014generative} that learn through a competitive process between two neural networks: a generator that creates synthetic data and a discriminator that distinguishes between real and generated samples. This adversarial training process enables GANs to generate ''realistic-looking'' predictions by transforming random latent vectors, along with conditioning information such as coarse-resolution predictors \citep{mirza2014conditional}, into stochastic outputs. The stochasticity arises from the random noise vector, allowing the model to produce multiple plausible solutions for the same input condition. However, one important limitation of GANs is that they are sensitive to the choice of training hyper-parameters \citep{rampal2025reliable} and can suffer from instabilities, hallucinations, and mode collapse \citep{radford2015unsupervised, arjovsky2017wasserstein}. To improve training stability, GAN-based algorithms commonly employ Wasserstein-distance based loss functions (WGANs) \citep{arjovsky2017wasserstein} with gradient penalty \citep{gulrajani2017improved}, as implemented in several recent climate downscaling studies \citep{leinonen2021stochastic, harris2022generative, price2022increasing, vosper2023deep, rampal2025reliable}. Other approaches, such as that in \citet{rampal2025reliable}, have incorporated intensity-constrained loss functions to significantly improve the accuracy of predicting extremes and other climatological statistics when downscaling daily precipitation.

Diffusion models and flow matching algorithms have recently emerged as alternative generative approaches to GANs for climate and weather downscaling applications \citep{addison2022machine, mardani2025residual,wetherell2026flow, addison2026machine}, which progressively denoise a sample from pure noise until it matches the data distribution. Although prediction or inference with diffusion and flow matching models often requires greater computational resources than GANs, diffusion models are more stable to train and have been shown to capture fine-scale spatial patterns with high fidelity, often better than GANs in unconditional image synthesis \citep{dhariwal2021diffusion}. Recent efforts to improve prediction fidelity have trained diffusion models to learn a residual correction objective, i.e., predicting a stochastic difference (residual) between deterministic model predictions (conditional mean) and the ground truth high-resolution fields \citep{mardani2025residual}. However, there are a limited number of studies that have impltemented and compared generative diffusion and flow matching models for climate downscaling \citep[e.g.,][]{addison2022machine, addison2026machine, liu2024downscaling, mardani2025residual, tomasi2025ldm, wetherell2026flow, legasa_regional_2026}. 

Despite progress in applying GANs, diffusion models and flow matching algorithms to climate downscaling, direct comparative studies between these approaches under consistent conditions remain limited, and the relative strengths and weaknesses of each method remain unclear. Recent studies by \citet{tomasi2025ldm}, \citet{schillinger2025enscale} and \citet{hobeichi2025applying} are among the few that have conducted such benchmarking, highlighting the need for further comparisons across different regions to assess the reliability and generalizability of these approaches, especially in regions with complex geography and diverse microclimates such as New Zealand \citep{sturman2006weather}. Building on earlier work that comprehensively evaluated GAN-based approaches for precipitation downscaling over New Zealand \citep{rampal2025reliable, rampal2024extrapolation}, this study extends the intercomparison to include diffusion models and flow matching algorithms across multiple configurations. These include residual and non-residual variants of each model type, as well as flow matching approaches using both Gaussian and Student-t noise priors. All methods are evaluated under consistent conditions using a comprehensive suite of metrics spanning spatial structure, distributional properties, ensemble dispersion and calibration, and climate change signals, providing practical guidance on model selection for applications ranging from climate projection to weather generation.

\section{Methodology}

We first describe the RCM simulations used for training and evaluation. Followed by descriptions of the generative AI models employed based on the residual correction framework \citep{mardani2025residual, rampal2025reliable}, and the evaluation metrics used in this study. 

\subsection{Training and Evaluation Data}\label{section:training-and-evaluation}

The downscaling models are trained and evaluated using 12 km dynamical downscaled simulations over New Zealand from the Conformal Cubic Atmospheric Model \citep[CCAM;][]{mcgregor2008updated}. These simulations have been comprehensively documented and evaluated in prior studies \citep{gibson2023high, gibson2025downscaled, gibson2024dynamical, campbell_comparison_2024, gibson2025downscaledtc} and have also been used to develop AI-based emulators \citep{lewis2025generative, rampal2024extrapolation, rampal2025reliable}. The RCM emulators developed in this study are trained on 137 years of CCAM output (1960–2096) driven by the ACCESS-CM2 GCM \citep{gibson2024dynamical}, where the period 2097-2099 is reserved for validation. The models are evaluated on four independent CCAM simulations driven by withheld GCMs (EC-Earth3, NorESM2-MM, AWI-CM-1-MR, and CNRM-CM6-1), totalling 560 years (140 years per simulation; 1960–2099) of ground-truth RCM output. The four evaluation GCMs span a wide range of warming rates \citep{rampal2024extrapolation, gibson2025downscaled}, ensuring that results reflect performance across a range of different climate conditions. Figures in the main text draw primarily from EC-Earth3; tabulated results are averaged across all four evaluation GCMs for a historical (1985--2014) and end-of-century future (2070--2099) period.

Both training and evaluation follow the ``perfect model framework", in which coarse-resolution prognostic variables from CCAM are used as predictors rather than fields directly from the driving GCM \citep[e.g.,][]{rampal2024enhancing, doury2023regional, van_der_meer_deep_2023}. This approach isolates the algorithms' extrapolation capabilities from discrepancies between RCM and GCM inputs. For consistency, we select the same set as \citet{rampal2025reliable}, namely: zonal wind speed ($U$) [m/s], meridional wind speed ($V$) [m/s], temperature ($T$) [K], and specific humidity ($Q$) [g/kg]. Each prognostic variable is selected from two simulated pressure levels in the atmosphere (500 and 850 hPa). To match typical GCM resolution \citep{maraun2016bias}, these fields have been re-gridded to a uniform 1.5\textdegree\, latitude-longitude grid ($\sim$130 km at 40\textdegree S) with conservative remapping. We also include topography (surface elevation) at a high resolution (12 km) as a static (time-invariant) predictor, as it has helped models learn location-specific biases (e.g., orographic rainfall) \citep{bailie2024quantile, rampal2025reliable}. All dynamic predictor fields are standardised based on the mean and standard deviation computed over the training dataset, as implemented in \citet{rasp2020weatherbench, rampal2022high, rampal2025reliable}, and topography is normalised to $[0,1]$ as implemented in \citet{rampal2025reliable}.

The target variable (i.e. downscaled output) is the daily accumulated precipitation [mm/day] from the high-resolution CCAM output ($\sim$12 km). We focus on daily predictor and target variables, as in previous work \citep{rampal2024extrapolation}, because it reduces the compute time and bandwidth, and allows our RCM emulator to be compatible with a broader range of GCMs (due to data availability). Because precipitation is significantly non-Gaussian \citep{renwick2009statistical}, we use a logarithmic transformation $ z_\text{pr}=\log_e(\text{pr}+1\text{ mm/day})$ for normalisation to reduce skewness, as implemented in previous work \citep{rampal2025reliable}. The spatial domain of the high-resolution fields is cropped to cover New Zealand and the adjacent ocean (165\textdegree E--184\textdegree W, 33\textdegree S--51\textdegree S), corresponding to a $168\times168$ grid with a resolution of $\sim$12 km. To provide sufficient context for downscaling at boundary edges, we take the coarse predictor fields on a slightly larger domain (151\textdegree E--188\textdegree W, 26\textdegree S--59\textdegree S), corresponding to a $21\times21$ grid with a resolution of 1.5\textdegree\, ($\sim$130 km at 40\textdegree S) after re-gridding. These domains match similar work on downscaling in New Zealand \citep{rampal2022high, bailie2024quantile, rampal2025reliable}.

\subsection{Overview of models}\label{section:intercomparison}

We compare eight models representing three families of widely used, state-of-the-art generative methods — GANs, diffusion models, and flow matching algorithms — each evaluated against a deterministic U-Net baseline (\autoref{tab:model_summary}). For the diffusion and flow matching approaches, we test both residual and non-residual (direct) variants. In residual configurations, a deterministic U-Net first predicts a conditional mean, and the generative model learns to correct the residuals from that prediction \citep{mardani2025residual}. In non-residual configurations, the generative model predicts the full output directly. The models and their variants are described in detail below.

\begin{table}[htbp]
  \centering
  \caption{\footnotesize Summary of models evaluated in this study. In model names, Res- denotes a residual configuration and Flow or FM denotes flow matching. Diffusion models use the Denoising Diffusion Probabilistic Model (DDPM) framework with Denoising Diffusion Implicit Model (DDIM) sampling for accelerated inference. Target indicates whether the model predicts the high-resolution field directly (direct) or predicts a correction to the U-Net output (residual). Abbreviations --- FiLM: Feature-wise Linear Modulation; CBAM: Convolutional Block Attention Module; AB2 PC: Adams--Bashforth 2-step Predictor--Corrector.}
  \label{tab:model_summary}
  \resizebox{\textwidth}{!}{
  \begin{tabular}{llclp{7cm}}
    \hline
    \textbf{Model} & \textbf{Framework} & \textbf{Target} & \textbf{Noise Prior} & \textbf{Key Features} \\
    \hline
    U-Net     & Deterministic CNN & Direct   & N/A         & MSE baseline; best-performing across residual configurations \\[4pt]
    ResGAN-v1 & Conditional GAN   & Residual & Latent      & Residual GAN; single-member MSE loss \\[4pt]
    ResGAN-v2 & Conditional GAN   & Residual & Latent      & As ResGAN-v1, with day-of-year predictor, ensemble-mean MSE, FiLM, self-attention, and CBAM \\[4pt]
    DDPM      & Diffusion (DDPM)  & Direct   & Gaussian    & ResGAN-v2 architecture with FiLM diffusion-timestep conditioning; non-residual; DDIM sampler (25 steps) \citep{song2020denoising} \\[4pt]
    ResDDPM   & Diffusion (DDPM)  & Residual & Gaussian    & As DDPM but residual \citep{mardani2025residual} \\[4pt]
    RCMFlow   & Flow Matching                & Direct   & Gaussian    & As DDPM but using flow matching; AB2 PC solver \citep{rampal2026cordexmlbench} \\[4pt]
    RCM-tFlow & Flow Matching                & Direct   & Student-$t$ & As RCMFlow but with heavy-tailed noise prior \citep{pandey2024heavy} \\[4pt]
    Res-tFM   & Flow Matching                & Residual & Student-$t$ & As RCM-tFlow but residual, following \citet{mardani2025residual} \\
    \hline
  \end{tabular}
  }
\end{table}
 
\subsection{Residual correction framework}\label{section:resdiual-correction-framework}
\begin{figure*}
    \centering
    \begin{adjustbox}{width=1.15\linewidth,center}
        \includegraphics{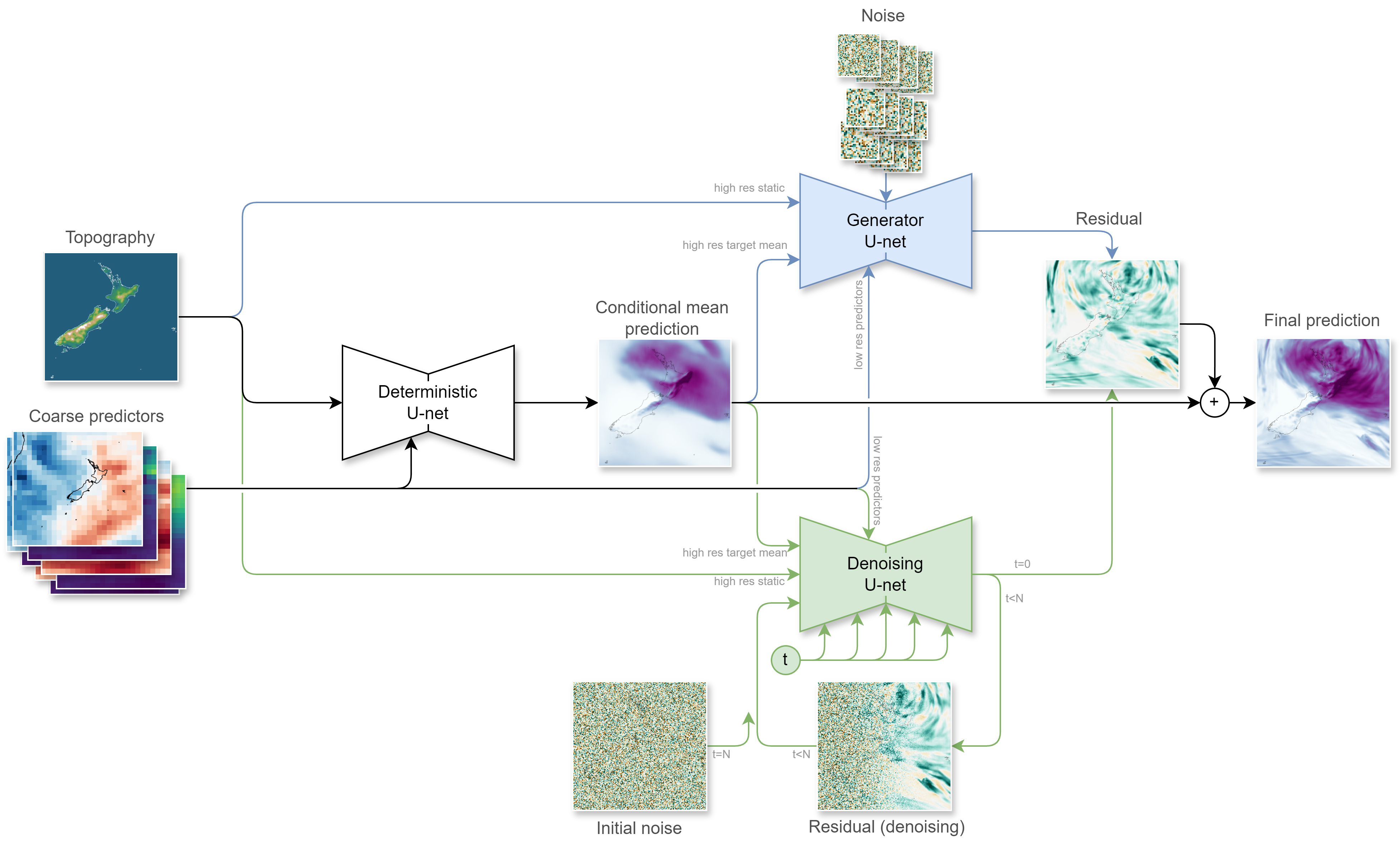}
    \end{adjustbox}
    \caption{\footnotesize Illustration of the residual correction framework applied to ResGAN (blue), and diffusion and flow matching models (green). In any given configuration, only one generative approach is used.  For the Denoising  U-Net schematic, the denoising illustration depicts the reverse diffusion process — a transition from Gaussian noise to the final prediction.}
    \label{fig:arch_full}
\end{figure*}

Four of the generative model configurations evaluated in this study use a residual correction framework \citep{mardani2025residual, rampal2025reliable}, illustrated in Fig.~\ref{fig:arch_full}. The residual configurations evaluated are ResDDPM, ResGAN-v1, ResGAN-v2, and Res-tFM. ResDDPM, ResGAN-v2, and Res-tFM share closely related architectures (described below), while ResGAN-v1 differs in both architecture and loss function (Table~\ref{tab:model_summary}), having been previously developed and validated for operational large-ensemble climate projection downscaling \citep{rampal2025reliable, rampal2025downscaling, rampal2024extrapolation}.

We define the residual correction procedure as the following: let $\mathbf x_{pred}$ represent the collection of input features, i.e. the coarse climate fields and the high-resolution topography, and $\mathbf y_{true}$ represent the high-resolution target field (precipitation). Under the residual correction framework, the target field is composed of a high-resolution conditional mean $\mathbf y_\text{det}$ and residual $\textbf r$ such that $\mathbf y_\text{det}+\mathbf r=\mathbf y_{true}$. The deterministic model is trained to predict the distribution of $\mathbf y$ given $\mathbf x$, and we denote the learned distribution as $\hat{\mathbf y}_\text{det}$. The generative model is then trained to model the residual distribution $\hat{\mathbf y}_\text{true}-\mathbf y_\text{det}=\mathbf r$ given the coarse climate fields $\mathbf x$ and the conditional mean $\hat{\mathbf y}_\text{det}$. We then sample from both models to produce a final high-resolution precipitation prediction $\hat{\mathbf{y}} = \hat{\mathbf{y}}_\text{det} + \hat{\mathbf{r}}$, where $\hat{\mathbf{r}}$ is the residual field predicted by the generative model and $\hat{\mathbf{y}}$ is the resulting full field.

\subsection{Deterministic baseline}\label{section:deterministic-baseline}

The deterministic baseline is a U-Net CNN, a fully convolutional encoder-decoder architecture \citep{ronneberger2015u} widely used in empirical downscaling \citep{rampal2024enhancing}, which learns a mapping from coarse predictor fields and high-resolution topography to the high-resolution precipitation target. The encoder progressively downsamples inputs through residual convolution blocks and pooling layers, and the decoder upsamples and refines these features to the high-resolution target. Two U-Net variants are used in this study; since the focus is on intercomparing generative approaches, we report results using the best-performing U-Net across all configurations. The U-Net paired with ResGAN-v1 follows \citet{rampal2025reliable}, using three downsampling levels with separate encoding paths for coarse predictor fields and topography, and no skip connections between encoder and decoder (Fig.~\ref{fig:arch_combined}a). All other U-Net configurations additionally incorporate: (1) Feature-wise Linear Modulation (FiLM), which adaptively scales and shifts feature maps conditioned on the day-of-year, allowing the model to modulate predictions based on seasonal context; (2) self-attention in the U-Net bottleneck, which captures long-range spatial dependencies across the full domain; and (3) the Convolutional Block Attention Module (CBAM), which applies lightweight channel- and spatial-attention gates to selectively emphasise informative features at each layer. These configurations share the same training hyperparameters. Note, we train a separate deterministic U-Net with each generative model configuration, as opposed to training a common backbone for the generative models. Since multiple U-Net configurations are trained, we use the best-performing one as a baseline; though most perform similarly given their identical hyperparameters and model selection, with the exception of the U-Net paired with ResGAN-v1.

\subsection{Residual GAN}\label{section:residual-GAN}

GANs, introduced by \citet{goodfellow2014generative}, consist of two competing networks: a generator that learns to produce synthetic data, and a discriminator that learns to distinguish generated data from real data. These two networks are trained with a minimax objective: the generator attempts to fool the discriminator, while the discriminator strives to correctly identify whether a sample is real or fake (i.e., generated). The aim is to have the networks converge where the generator reproduces the true data distribution and the discriminator can no longer distinguish generated samples from real data. An important extension is the conditional GAN (hereon simply referred to as GAN) \citep{mirza2014conditional}, where additional conditioning information (e.g., class labels, or a data vector) is fed into both the generator and the discriminator. Rather than just generating samples from the data distribution, images can be generated based on class labels or lower-resolution images, for example. Conditioning the generator is necessary for using GANs in climate downscaling, as we want the generator to produce high-resolution fields given a low-resolution input.

The residual conditional GAN model learns to generate a high-resolution precipitation field by correcting the output from the deterministic U-Net. We denote the deterministic prediction by $\hat{\mathbf y}_\text{det}$, which, along with the conditioning data $\mathbf x$ (coarse predictor fields and static topography) and randomly sampled Gaussian noise vector $\mathbf z$, the GAN generates the precipitation residual $\mathbf r_\text{GAN} = G_\theta(\mathbf r_\text{GAN}; \hat{\mathbf y}_\text{det}, \mathbf x, \mathbf z)$. The final high-resolution precipitation prediction is then obtained by adding the residual to the deterministic prediction: $\hat{\mathbf y} = \hat{\mathbf y}_\text{det} + \mathbf r_\text{GAN}$. The purpose of $\mathbf r_\text{GAN}$ is to represent the high-frequency details and adjustments needed to transform the smooth deterministic U-Net prediction into a realistic sample \citep{rampal2025reliable}.

\begin{figure}
    \centering
    \textbf{(a)}
    
    \vspace{0.5em}
    
    \begin{adjustbox}{width=0.7\linewidth,center}
        \includegraphics{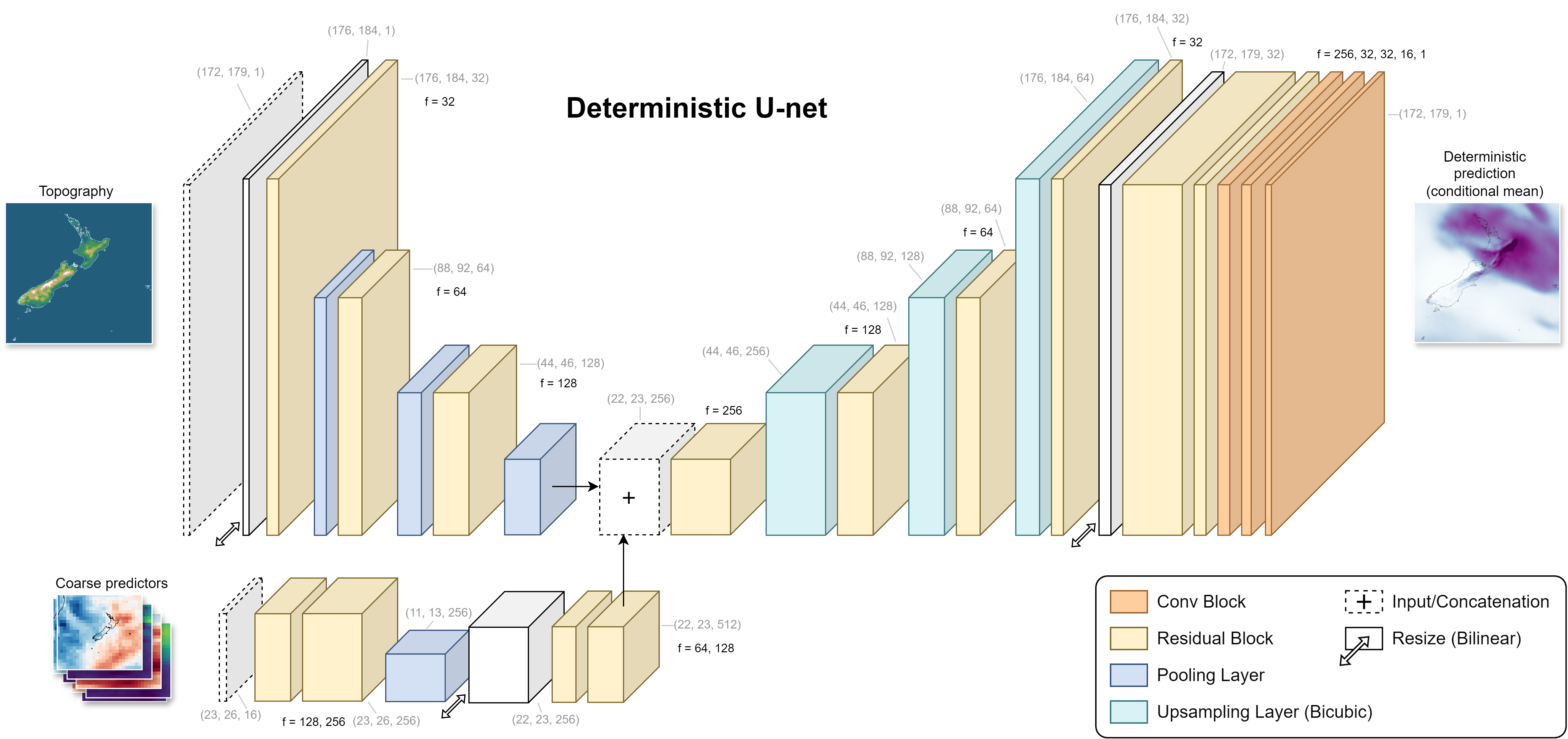}
    \end{adjustbox}
    
    \vspace{1em}
    
    \textbf{(b)}
    
    \vspace{0.5em}
    
    \begin{adjustbox}{width=0.7\linewidth,center}
        \includegraphics{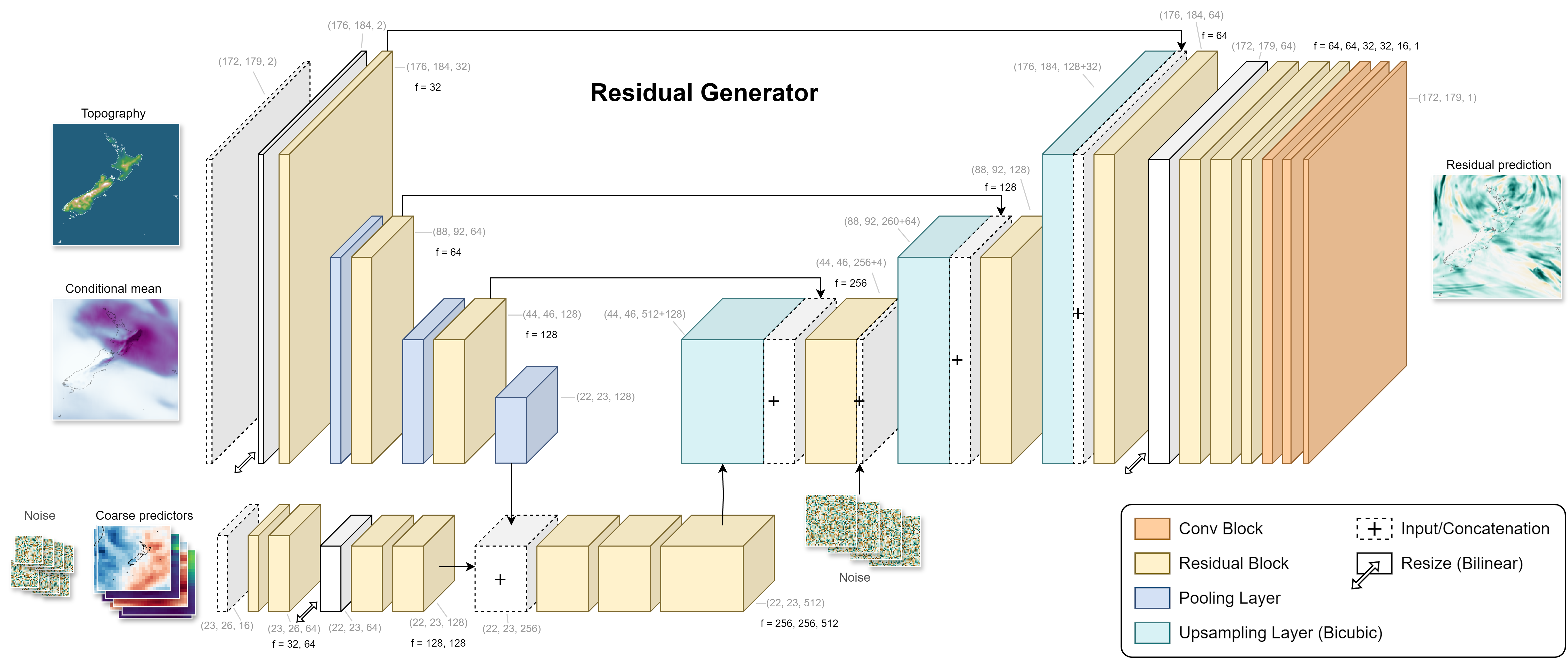}
    \end{adjustbox}
    
    \vspace{1em}
    
    \textbf{(c)}
    
    \vspace{0.5em}
    
    \begin{adjustbox}{width=0.7\linewidth,center}
        \includegraphics{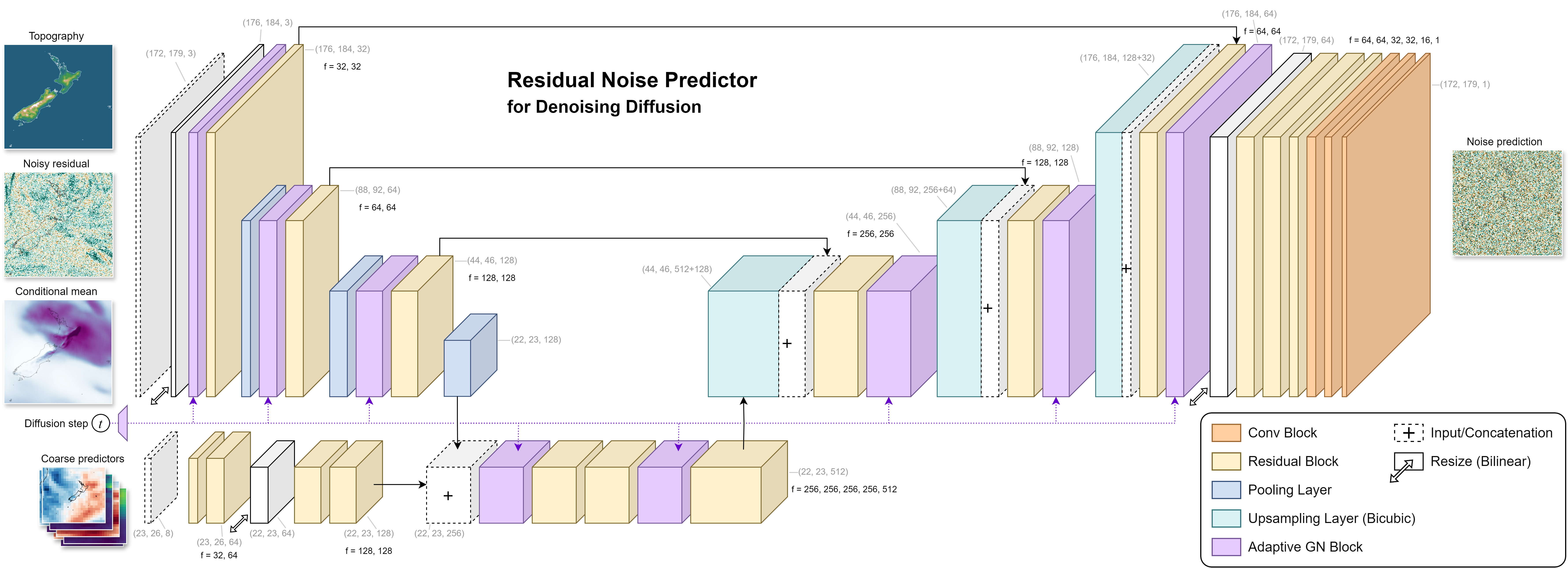}
    \end{adjustbox}
    
    \vspace{1em}
    
    \textbf{(d)}
    
    \vspace{0.5em}
    
    \begin{adjustbox}{width=0.65\linewidth,center}
        \includegraphics{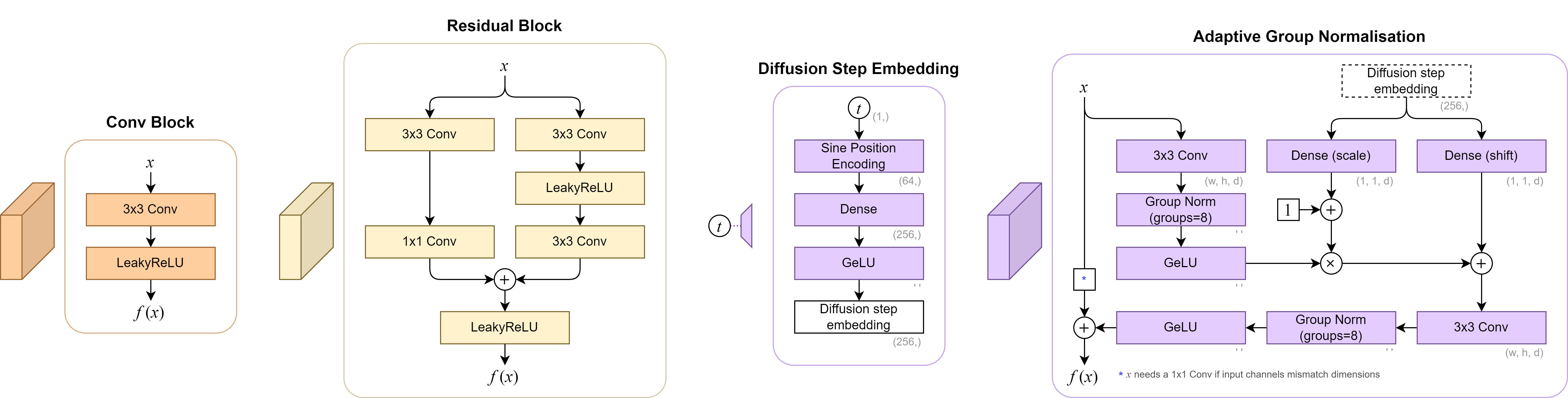}
    \end{adjustbox}
    \caption{\footnotesize \textbf{a)} Architecture of the deterministic U-net used for downscaling coarse-resolution predictors into the high-resolution precipitation field. The network takes a static high-resolution topography field and concatenated coarse predictors as inputs $\mathbf{x}$, which are processed through successive encoding layers (residual and pooling blocks) before being passed through a bottleneck, where the encoded topography and predictor features are concatenated. The decoder progressively upsamples the features using bicubic interpolation and residual blocks, and the final convolutional layers generate the deterministic prediction $\hat{\mathbf{y}}_\text{det}$ for the high-resolution precipitation field. \textbf{b)} Architecture of the ResGAN-v1 for predicting a residual to correct $\hat{\mathbf{y}}_\text{det}$. Similar to (a), but takes additional inputs: $\hat{\mathbf{y}}_\text{det}$ as the conditional mean and noise vectors. High-resolution inputs are processed through encoding layers, while coarse fields and noise pass through residual blocks without pooling. After concatenation in the bottleneck, features are upsampled with skip connections from the encoder. Additional noise is injected after the first upsample. Outputs residual prediction $\hat{\mathbf{r}}_\text{GAN}$ for small-scale corrections ($\hat{\mathbf{y}}_\text{GAN}=\hat{\mathbf{r}}_\text{GAN}+\hat{\mathbf{y}}_\text{det}$). \textbf{c)} Architecture of ResDDPM for correcting $\hat{\mathbf{y}}_\text{det}$. Takes diffusion step $t$ and noisy residual $\hat{\mathbf{r}}_t$, conditioned by $\hat{\mathbf{y}}_\text{det}$ and inputs $\mathbf{x}$. Similar encoder-decoder structure as (a), but with adaptive group normalization blocks. Outputs noise prediction $\hat{\boldsymbol{\epsilon}}$ to obtain denoised residual $\hat{\mathbf{r}}_0\gets \hat{\mathbf{r}}_t-\hat{\boldsymbol{\epsilon}}$. (d) Definitions of network components used in (a), (b), and (c).}
    \label{fig:arch_combined}
\end{figure}

In this study, two GAN variants are evaluated: ResGAN-v1 and ResGAN-v2, which differ in their U-Net architecture, layer choices, and loss function. ResGAN-v1 was developed in \citet{rampal2025downscaling} with a focus on reliably capturing climate change signals, and is trained with a single-member MSE loss. Both generators are based on the U-Net architecture described previously. As illustrated in Fig.~\ref{fig:arch_combined}b, the generator $G_\theta$ follows \citet{rampal2025reliable}, with three key differences from the deterministic U-Net: (1) the input channels include the deterministic prediction $\hat{\mathbf y}_\text{det}$ alongside the coarse predictor fields and topography $\mathbf x$; (2) upsampling blocks receive encoder outputs via skip connections, as is standard in U-Net architectures \citep{ronneberger2015u}; and (3) two noise vectors are injected into the generator — the first concatenated with the bottleneck features and the second with the output of the first upsampling block. ResGAN-v2 extends this design with several additions aimed at improving ensemble dispersion and general skill: an ensemble-mean MSE loss, where the mean is computed across ensemble members prior to the loss calculation \citep{harris2022generative}; self-attention in the bottleneck following \citet{mardani2025residual}; FiLM conditioning on day-of-year; and CBAM attention layers. Both variants share an identical discriminator architecture.

The discriminator $D_\theta$, based on \citet{rampal2025reliable}, is a CNN that estimates the probability that a precipitation field $\mathbf y$ is a real sample rather than a generator output. The GAN is trained with an adversarial objective \citep{goodfellow2014generative}: the discriminator is trained to maximise its accuracy in identifying real versus generated fields using Wasserstein critic loss with gradient penalty \citep{arjovsky2017wasserstein, gulrajani2017improved}, while the generator is trained to minimise the content loss while maximising the adversarial loss. In addition to the content loss, an intensity constraint is applied to penalise inaccurate precipitation extremes \citep{rampal2025reliable}. The loss functions use to train the generator and discriminators are shown below.

$$\mathbf r_\text{true}= \hat{\mathbf y}_\text{true} - \mathbf y_\text{det}
\,,\qquad \hat{\mathbf y}_\text{GAN}=\hat{\mathbf y}_\text{det} + \hat{\mathbf r}_\text{GAN}$$

$$
\hat{\mathbf r}_\text{lerp} = {\mathbf r}_\text{true} + \alpha(\hat{\mathbf r}_\text{GAN} - {\mathbf r}_\text{true})
\,,\quad \alpha\sim\mathcal N(0,1)$$

$$
\mathcal L_\text{D}=
\underbrace{
    \overline{D_\theta(\hat{\mathbf r}_\text{GAN})} - \overline{D_\theta(\mathbf r_\text{true})}
}_{\text{critic loss}}
+ \underbrace{
    \gamma
    \overline{(\lVert \nabla_{\hat{\mathbf r}_\text{lerp}}\,D_\theta(\hat{\mathbf r}_\text{lerp}) \rVert_2-1)^2}
}_{\text{gradient penalty}}
$$

$$
\mathcal L_\text{G}=
\underbrace{
    \text{MSE}(\mathbf r_\text{true},\hat{\mathbf r}_\text{GAN})
}_{\text{content loss}}
- \underbrace{
    \lambda_\text{adv}\overline{D_\theta(\hat{\mathbf r}_\text{GAN})}
}_{\text{adversarial loss}}
+ \underbrace{
    \text{MSE}(\mathbf y_\text{true}^\text{max},\hat{\mathbf y}_\text{GAN}^\text{max})
}_{\text{intensity constraint}}
$$.

As previously discussed in \citet{rampal2025reliable}, the GAN is trained to predict residuals rather than absolute precipitation values. Following the WGAN with gradient penalty (WGAN-GP) framework \citep{gulrajani_improved_2017}, a gradient penalty term is applied during training to enforce Lipschitz continuity of the critic (discriminator). This gradient penalty ($(\lVert \nabla_{\hat{\mathbf r}_\text{lerp}}\,D_\theta(\hat{\mathbf r}_\text{lerp}) \rVert_2-1)^2$) is computed on interpolated samples that combine real and generated residuals ($\hat{\mathbf{r}}_\text{lerp}$),
where $\alpha$ is randomly sampled from a uniform distribution. Here, $D_{\theta}$ denotes the critic and $\gamma$ is the penalty coefficient. The generator loss consists of three components as outlined in \citet{rampal2025reliable}: the adversarial loss (critic score), a mean squared error term to ensure fidelity to the deterministic baseline, and a maximum-intensity loss to preserve extreme precipitation events.

The GAN models were trained for 200 epochs, using the hyperparameters described in \citet{rampal2025reliable}. Specifically, we use the Adam optimizer \citep{kingma2014adam} with a learning rate of 0.00015 and momentum $\beta_1=0.5,\beta_2=0.9$. The discriminator learning rate is fixed while the deterministic U-net and generator learning rate decay by 0.995 every 1000 steps. The discriminator is trained for three steps for each generator training step with a gradient penalty ($\gamma$) weight of 10. The generator loss has an adversarial loss weight of 0.01 and an intensity constraint weight of 1.

\subsection{Residual and non-residual diffusion}\label{section:residual-diffusion}

The second generative approach evaluated is a diffusion model based on the Denoising Diffusion Probabilistic Model (DDPM) framework \citep{ho2020denoising}, tested in both residual (ResDDPM) and non-residual (hereafter referred to as DDPM) configurations. In the residual configuration, the diffusion model learns to predict the precipitation residual from the deterministic U-Net output; in the non-residual configuration, it predicts the full high-resolution field directly. Diffusion models generate synthetic data through a two-stage process, unified here under the stochastic differential equation (SDE) framework \citep{song2020score}. In the forward process, Gaussian noise is progressively added to training samples over $T$ timesteps — defined as a Markov chain $q(x_t|x_{t-1})$ — until the data is indistinguishable from pure Gaussian noise, $\mathcal{N}(0,1)$. The reverse process inverts this by solving a reverse-time SDE driven by a learned score function, gradually denoising samples to recover realistic outputs \citep{sohldickstein2015deep, ho2020denoising}. Alternatively, samples can be generated deterministically via a probability-flow ordinary differential equation (ODE), which enables likelihood evaluation and the use of higher-order ODE solvers \citep{song2020score}. A neural network then learns the reverse process $p_\theta(\mathbf{x}_{t-1}|\mathbf{x}_t, \mathbf{X})$, which denoises the signal one step at a time conditioned on the coarse-resolution predictor fields $\mathbf{X}$. This conditional formulation is a key distinction from unconditional diffusion models \citep{ho2020denoising}, where the reverse process $p_\theta(\mathbf{x}_{t-1}|\mathbf{x}_t)$ operates without any external conditioning signal and generates samples solely from the learned data distribution. Here, conditioning on $\mathbf{X}$ at each denoising step steers the generated field towards a plausible high-resolution precipitation field given the large-scale atmospheric state. The training objective is simplified by parameterising the network to predict the Gaussian noise added at each timestep, reducing the loss to a mean-squared error between predicted and true noise. During inference or prediction, the model begins from random Gaussian noise $\mathbf{x}_T \sim \mathcal{N}(\mathbf{0}, \mathbf{I})$ and applies the learned reverse transitions $p_\theta(\mathbf{x}_{t-1}|\mathbf{x}_t, \mathbf{X})$ for $t = T, \dots, 1$ to produce a realistic precipitation residual sample $\hat{\mathbf{x}}_0$.



\subsubsection{Forward diffusion (noising)}
The forward diffusion process transforms the residual precipitation distribution $q(\mathbf{r}_0)$ into a standard Gaussian $\mathcal{N}(\mathbf{0}, \mathbf{I})$ \citep{sohldickstein2015deep, ho2020denoising} via a Markov chain:
$$q(\mathbf{r}_{1:T}) = q(\mathbf{r}_0)\prod_{t=1}^T q(\mathbf{r}_t|\mathbf{r}_{t-1}),$$
where each noising step adds variance $\beta_t \in (0,1)$:
$$q(\mathbf{r}_t|\mathbf{r}_{t-1}) = \mathcal{N}(\mathbf{r}_t;\, \sqrt{1-\beta_t}\,\mathbf{r}_{t-1},\, \beta_t\mathbf{I}).$$
A key property of the forward process is a closed-form solution for any arbitrary timestep $t$:
$$q(\mathbf{r}_t|\mathbf{r}_0) = \mathcal{N}(\mathbf{r}_t;\, \sqrt{\bar{\alpha}_t}\,\mathbf{r}_0,\, (1-\bar{\alpha}_t)\mathbf{I}),$$
where $\alpha_t := 1-\beta_t$ and $\bar{\alpha}_t := \prod_{i=1}^t \alpha_i$ \citep{ho2020denoising}, allowing a clean residual $\mathbf{r}_0$ to be noised to $\mathbf{r}_t$ in a single step:
$$\mathbf{r}_t(\mathbf{r}_0, \boldsymbol{\epsilon}) = \sqrt{\bar{\alpha}_t}\,\mathbf{r}_0 + \sqrt{1-\bar{\alpha}_t}\,\boldsymbol{\epsilon}, \quad \boldsymbol{\epsilon} \sim \mathcal{N}(\mathbf{0}, \mathbf{I}).$$
Note that the forward process is unconditional — it depends only on $\mathbf{r}_0$ and does not use the coarse predictor fields $\mathbf{x}$ or the deterministic prediction $\hat{\mathbf{y}}_\text{det}$. Conditioning is introduced exclusively in the reverse process described below.

\subsubsection{Reverse diffusion (denoising)}
Unlike the forward process, the reverse process is conditional: given the coarse predictor fields $\mathbf{x}$ and the deterministic U-Net prediction $\hat{\mathbf{y}}_\text{det}$, the model learns to denoise $\mathbf{r}_t$ back towards a realistic precipitation residual. The conditional reverse process is:
$$p_\theta(\mathbf{r}_{t-1}|\mathbf{r}_t, \mathbf{x}, \hat{\mathbf{y}}_\text{det}) = \mathcal{N}(\mathbf{r}_{t-1};\, \mu_\theta(\mathbf{r}_t, t, \mathbf{x}, \hat{\mathbf{y}}_\text{det}),\, \beta_t\mathbf{I}).$$
A U-Net-based network is trained to predict the noise $\epsilon_\theta$ (Fig.~\ref{fig:arch_combined}), with the training loss reduced to the MSE between true and predicted noise:
$$\mathcal{L}_\theta = \left\|\boldsymbol{\epsilon} - \epsilon_\theta\!\left(\sqrt{\bar{\alpha}_t}\,\mathbf{r}_\text{true} + \sqrt{1-\bar{\alpha}_t}\,\boldsymbol{\epsilon};\; t, \mathbf{x}, \hat{\mathbf{y}}_\text{det}\right)\right\|^2.$$
The predicted noise is used to recover the conditional mean $\mu_\theta$, from which $\hat{\mathbf{r}}_{t-1}$ is sampled progressively to obtain a denoised residual estimate $\hat{\mathbf{r}}_0$:
$$\hat{\mathbf{r}}_{t-1} = \mu_\theta(\mathbf{r}_t, t, \mathbf{x}, \hat{\mathbf{y}}_\text{det}) = \frac{1}{\sqrt{\alpha_t}}\left(\mathbf{r}_t - \frac{\beta_t}{\sqrt{1-\bar{\alpha}_t}}\,\epsilon_\theta(\mathbf{r}_t, t, \mathbf{x}, \hat{\mathbf{y}}_\text{det})\right).$$

\subsubsection{Denoiser network}

The denoiser network $\epsilon_\theta$ is implemented as a U-Net sharing the same base architecture as ResGAN-v2, modified in two ways: the latent noise inputs are removed, and a diffusion timestep embedding is added. At each training step, the input to the network is a noisy residual $\mathbf{r}_t = \mathbf{r}_t(\mathbf{y}_\text{true} - \hat{\mathbf{y}}_\text{det}, \boldsymbol{\epsilon})$ concatenated with the deterministic prediction $\hat{\mathbf{y}}_\text{det}$ and static topography. The coarse predictor fields are processed through residual blocks before being concatenated with the encoder output, and skip connections are used between encoder and decoder levels (Fig.~\ref{fig:arch_combined}c). The diffusion timestep is encoded as a sinusoidal embedding (256 channels) and injected after the first activation via FiLM modulation:
$$h'_{cij} = h_{cij}(1 + \gamma_c(t)) + \beta_c(t),$$
where $\gamma, \beta \in \mathbb{R}^C$ are learned linear projections of the timestep embedding. The decoder upsamples via bicubic interpolation, and the network outputs the predicted noise $\hat{\boldsymbol{\epsilon}}$, trained to minimise $\|\boldsymbol{\epsilon} - \hat{\boldsymbol{\epsilon}}\|^2$.

\begin{figure}
    \centering
    \begin{adjustbox}{width=1.45\linewidth,center}
        \includegraphics[trim=0 45cm 0 0, clip]{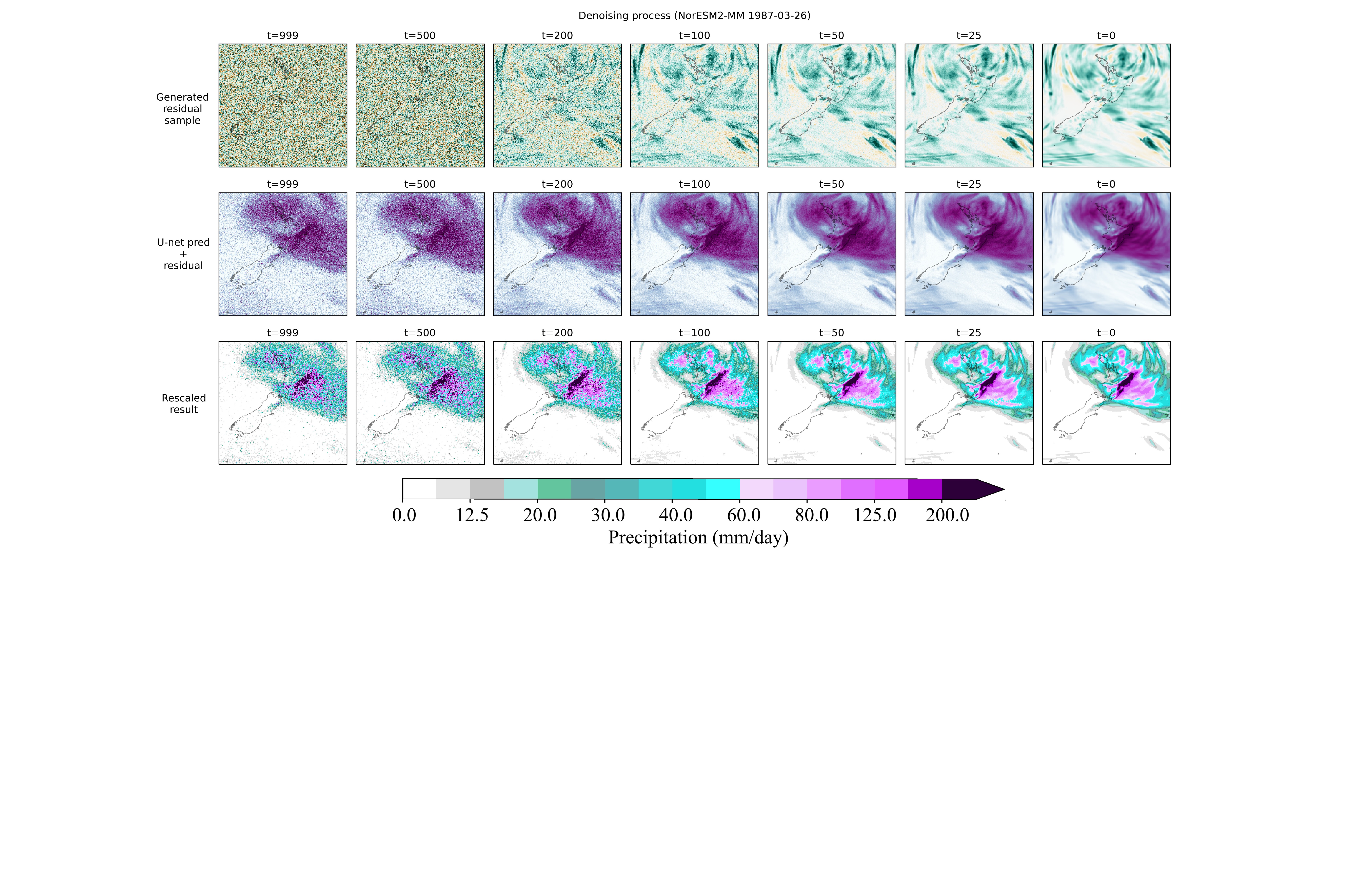}
    \end{adjustbox}
    \caption{\footnotesize A representation of the reverse diffusion process from time T=999 to t=0. The top row shows residuals (relative to the deterministic baseline) evolving from white noise at T=999 to the final residual prediction at t=0. The middle row shows the residuals added to the deterministic baseline in logarithmic space (because precipitation is log-normalized, as described in Section \ref{section:training-and-evaluation}). The bottom row shows the final precipitation field after reversing the normalization and taking the exponential of the values.}
    \label{fig:denoise_process}
\end{figure}

During inference, we use a linear noise schedule with $\beta_1 = 10^{-4}$ to $\beta_T = 0.02$ and DDIM sampling \citep{song2020denoising}, which adopts a non-Markovian reverse process to enable deterministic generation with far fewer steps than standard probabilistic sampling. For the residual configuration, the final high-resolution prediction is obtained as $\hat{\mathbf{y}}_0 = \hat{\mathbf{y}}_\text{det} + \hat{\mathbf{r}}_0$. We use 25 denoising steps, which provides a good balance between inference speed and performance; sensitivity experiments with 50 and 100 steps showed comparable results, with further analysis provided in Supplementary Figure S3. An illustration of the full denoising process, from Gaussian noise to the final prediction, is shown in Figure~\ref{fig:denoise_process}.

\subsubsection{Hyperparameters}

Models were trained for 200 epochs using an exponential moving average (EMA) of the weights with decay rate $\beta = 0.999$, which stabilises training by smoothing noisy parameter updates \citep{kingma2014adam, dhariwal2021diffusion}:
$$\bar{\theta}' \gets \beta\,\bar{\theta} + (1-\beta)\theta,$$
with $\bar{\theta}_0 = 0$. EMA weights are used at inference. Sensitivity experiments with $\beta = 0.5$ and without EMA produced substantially degraded results and are excluded. All diffusion models were trained with the same Adam optimiser and learning rate schedule as the GAN. Unlike the GAN, no additional loss constraints were applied to the denoiser, as we aimed to assess how a standard diffusion setup performs relative to a carefully tuned GAN. However, applying such constraints to the diffusion model proved non-trivial, as the network predicts noise rather than precipitation directly, and did not yield skillful results in practice.

\subsection{Flow matching}\label{section:flow-matching}

Flow matching \citep{lipman2022flow} is somewhat similar to diffusion models but, rather than learning to denoise, it learns to predict a velocity field that transports samples along straight-line paths from a base distribution to the data distribution. We evaluate three flow matching configurations: two non-residual variants with Gaussian (RCMFlow) and Student-$t$ (RCM-tFlow) noise priors, and one residual variant with a Student-$t$ prior (Res-tFM).

\subsubsection{Forward process and training objective}

The forward process defines a straight-line interpolant between a noise sample $\mathbf{x}_0$ drawn from the base distribution and the target $\mathbf{x}_1$ (the high-resolution precipitation field, or residual $\mathbf{y}_\text{true} - \hat{\mathbf{y}}_\text{det}$ in the residual configuration):
$$\mathbf{x}_t = (1-t)\,\mathbf{x}_0 + t\,\mathbf{x}_1, \quad t \in [0,1],$$
with a constant target velocity $\mathbf{v}^* = \mathbf{x}_1 - \mathbf{x}_0$. A neural network $v_\theta(\mathbf{x}_t, t, \mathbf{x})$ is trained to predict this velocity via MSE:
$$\mathcal{L} = \mathbb{E}_{t,\mathbf{x}_0,\mathbf{x}_1}\left[\left\|v_\theta(\mathbf{x}_t, t, \mathbf{x}) - (\mathbf{x}_1 - \mathbf{x}_0)\right\|^2\right].$$

In the standard Gaussian variant (RCMFlow), the base distribution is $\mathbf{x}_0 \sim \mathcal{N}(\mathbf{0}, \sigma_z^2\mathbf{I})$, where $\sigma_z$ is an adaptive noise scale set to the RMSE of the deterministic U-Net. Since log-transformed precipitation residuals exhibit heavy-tailed behaviour. Following \citet{pandey2024heavy}, we adopt a heavy-tailed variant that replaces the Gaussian prior with a Student-$t$ distribution, sampled as:
$$\mathbf{x}_0 = \sigma_z\,\frac{\mathbf{z}}{\sqrt{v/\nu}}, \quad \mathbf{z} \sim \mathcal{N}(\mathbf{0},\mathbf{I}),\quad v \sim \chi^2(\nu),$$
where $\nu = 5$ controls the tail thickness. The interpolation path and training objective are otherwise identical to the Gaussian case.

\subsubsection{Inference}

During inference, the precipitation field is recovered by integrating the learned ODE from $t=0$ to $t=1$:
$$\frac{\mathrm{d}\mathbf{x}}{\mathrm{d}t} = v_\theta(\mathbf{x}(t), t, \mathbf{x}).$$
Rather than simple Euler integration, we use an Adams--Bashforth 2-step Predictor--Corrector (AB2 PC) solver, which incorporates the velocity from the previous timestep to achieve second-order accuracy:
$$\mathbf{x}_{n+1} = \mathbf{x}_n + \frac{\Delta t}{2}\left[3\,v_\theta(\mathbf{x}_n, t_n, \mathbf{x}) - v_\theta(\mathbf{x}_{n-1}, t_{n-1}, \mathbf{x})\right].$$
This higher-order integration traces the deterministic flow trajectory more accurately than Euler. 

\subsubsection{Architecture and hyperparameters}

The velocity network $v_\theta$ shares the same architecture as the DDPM denoiser (Section~\ref{section:residual-diffusion}), with the diffusion timestep embedding replaced by a flow-time embedding. All flow matching models were trained for 250 epochs using the same Adam optimiser, learning rate schedule, and EMA weight averaging ($\beta = 0.999$) as the diffusion models, with EMA weights used at inference. As with the diffusion models, no additional loss constraints beyond the MSE velocity objective were applied.

\subsection{Evaluation metrics}\label{section:evaluation-metrics}
As described in Section~\ref{section:training-and-evaluation}, models are trained on the CCAM-downscaled ACCESS-CM2 simulation (1970--2099), with the four remaining GCM-driven simulations reserved for testing. Evaluation focuses on a historical period (1985--2014) and an end-of-century future period (2070--2099), with tabulated scores averaged across both periods and all four evaluation GCMs. For these evaluations, a 5-member ensemble is generated for each day for all generative approaches, with metrics computed per member and then averaged across members to improve robustness. In Section~\ref{section:sampling-sensitivity}, where ensemble dispersion and calibration are the focus, a 10-member ensemble is used for a case study over the year 2097 (excluded from training; 365 days).

Emulator skill is assessed using metrics spanning climatological seasonal means (DJF and JJA), wet extremes (Rx1Day; annual maximum 1-day precipitation), dry extremes (CDD; consecutive dry days), precipitation intensity distributions (LHD; logarithmic histogram distance), and spatial structure (RALSD; radially averaged logarithmic spectral distance). Climatological means of seasonal precipitation, wet extremes, and dry extremes are computed following \citet{rampal2025reliable}. The metrics LHD and RALSD are important for assessing whether the model can represent the full range of precipitation intensities and resolve fine-scale precipitation  \citep{rampal2025reliable}. Unlike traditional metrics (e.g., KL-divergence or perkins skill score), LHD weights errors in low-probability, high-intensity precipitation events more heavily. Additionally, RALSD weights errors at small scales (high wavenumbers, low energy) equally to errors at large scales (low wavenumbers, high energy), ensuring that fine-scale features are adequately evaluated.

For computing LHD, we construct a one-dimensional histogram $\gamma$ using all grid points and timesteps where daily precipitation exceeds 1 mm/day. The LHD is defined as:
$$\mathrm{LHD(dB)}=\sqrt{\frac{1}N\sum_{i=1}^N\left(10\log\frac{{\gamma_{\mathrm{true}}}_{i}}{{\gamma_{\mathrm{pred}}}_{i}}\right)^2},$$
which measures the logarithmic distance between the predicted histogram ${\gamma_{\mathrm{pred}}}_{i}$ and the ground truth histogram ${\gamma_{\mathrm{true}}}_{i}$ (from CCAM) across $i$ bins, excluding bins where ${\gamma_{\mathrm{true}}}_{i}$ has fewer than 10 counts. Following \citet{rampal2025reliable}, we use 53 evenly spaced bins spanning 1–1,050 mm/day with 20 mm spacing.
For the RALSD, we follow the approach of \citet{harris2022generative} and \citet{rampal2025reliable} to measure how well model predictions represent the true power spectral density (PSD), which indicates skill at resolving fine-scale precipitation structures. We first compute a Fourier transform on all predictions across the domain, then radially integrate over all angular directions using binning to form a one-dimensional power spectrum. The RALSD is defined as:

$$\mathrm{RALSD(dB)}=\sqrt{\frac 1N\sum_i\left(10\log\frac{
{{}\hat F_\mathrm{true}}_i
}{{{}\hat F_\mathrm{pred}}_i
}\right)^2},$$
which measures the logarithmic distance between the predicted PSD ${{}\hat{F}_\mathrm{pred}}_i$ and ground truth PSD ${{}\hat{F}_\mathrm{true}}_i$ (from CCAM). Following \citet{rampal2025reliable}, we use 26 bins between 0 and 0.5 with 0.02 spacing and normalize precipitation for each day before computing the Fourier transform. Similar to \citet{rampal2025reliable}, we use only the 200 rainiest days on average to compute the RALSD; however, we obtain similar results using all days.

We also evaluate the emulators' ability to capture climate change responses by computing the percentage change in precipitation between the historical period (1986–2005) and the future period (2080–2099). We focus on DJF mean precipitation and Rx1Day as representative metrics, spanning mean precipitation changes and climate change signals in extremes — the latter being inherently noisy and particularly challenging to emulate \citep{gibson2025downscaled}. Accurate reproduction of the climate change signal is essential for ML-based downscaling, yet this evaluation is frequently omitted from downscaling benchmarks \citep{rampal2024enhancing, kendon_potential_2025, hobeichi2025applying}.

\section{Results}

We first begin by characterising the sensitivity of a flow matching model to ODE solver
choice and number of sampling steps
(Section~\ref{section:sampling-sensitivity}), before examining model behaviour
during a high-impact weather event to illustrate differences in spatial
structure and ensemble spread (Section~\ref{section:intensity-structure}).
We then assess systematic performance across the full evaluation period using a
suite of climatological metrics (Section~\ref{section:climatology-metrics}),
and conclude by evaluating each model's ability to reproduce climate change
signals (Section~\ref{section:climate-change}).

\subsection{Effect of solver selection and sampling steps}
\label{section:sampling-sensitivity}

An important practical consideration for flow matching and diffusion models is the number of sampling steps — or function evaluations — required to solve the ODE and achieve satisfactory prediction quality, a quantity that can in principle be chosen freely.
Early implementations with diffusion required ${\sim}1000$~steps for high-fidelity
image output \citep{ho2020denoising}; however, deterministic samplers that bypass
the Markovian assumption can dramatically reduce this cost with little loss in
quality \citep{song2020denoising}.

\begin{figure}
    \centering
    \includegraphics[width=1.15\linewidth]{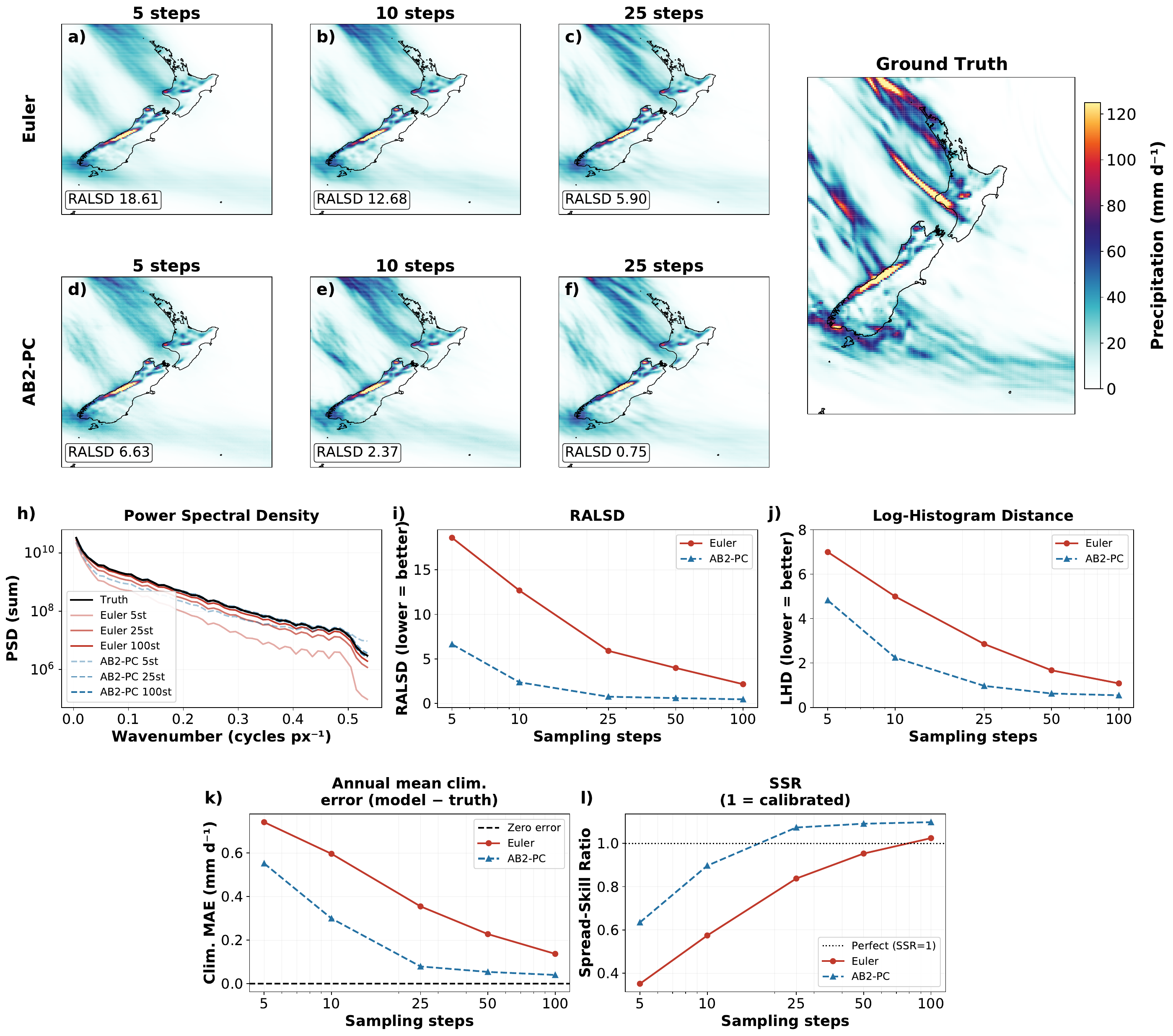}
    \caption{\footnotesize Sensitivity of Res-tFM flow matching model to solver choice
    (Euler vs.\ AB2~PC) and number of sampling steps, evaluated on a
    10-member ensemble for one year (2097).
    (a--f)~Example precipitation fields at 5, 10, and 25~steps for each
    solver, with RALSD scores inset.
    (g)~Ground truth CCAM field.
    (h)~Power spectral density curves for all configurations.
    (i)~RALSD, (j)~log-histogram distance (LHD),
    (k)~annual mean climatological MAE, and
    (l)~spread-skill ratio (SSR; perfect calibration $= 1$), each as a
    function of sampling steps.
    Lower values indicate better performance for RALSD, LHD, and
    climatological MAE.}
    \label{fig:sampling_speed}
\end{figure}

To study sensitivity to solver choice and number of sampling steps, we generate a 10-member ensemble over one year (2097) at 5, 10, 25, 50, and 100 steps. For flow matching models, we compare a first-order Euler solver against the Adams--Bashforth 2-step Predictor--Corrector (AB2-PC); for diffusion models, we use DDIM. Notably, AB2-PC requires no additional network evaluations relative to Euler: it reuses the velocity field from the previous timestep, falling back to a standard Euler step only at the first iteration. Figure~\ref{fig:sampling_speed} illustrates the effect of solver choice and step count for Res-tFM.

The example spatial fields (Figs.~\ref{fig:sampling_speed}a--f) show that AB2-PC recovers substantially more fine-scale structure than Euler at any given step count, which is also reflected quantitatively across all metrics. In terms of spatial structure, AB2-PC with just 25~steps (RALSD $= 0.75$) outperforms Euler with 100~steps (RALSD $= 1.17$; Figs.~\ref{fig:sampling_speed}h--i). Similar improvements are evident in the precipitation intensity distribution (LHD; Fig.~\ref{fig:sampling_speed}j) and annual mean climatological error (Fig.~\ref{fig:sampling_speed}k). Ensemble calibration is measured by the spread-skill ratio \citep[SSR;][]{leutbecher2008ensemble}, where a well-calibrated ensemble has a spread — the standard deviation across members averaged over time — that matches the ensemble mean error. AB2-PC achieves well-calibrated spread (SSR $\approx 1$) at just 25~steps, while Euler requires more than 100~steps to reach comparable calibration (Fig.~\ref{fig:sampling_speed}l). This result holds across all other flow matching configurations. For diffusion models, DDIM sampling \citep{song2020denoising} achieves comparably strong performance after 25~steps (Supplementary Figures S1-S3). Based on these results, we adopt 25~steps with AB2-PC for all flow matching models and DDIM with 25~steps for all diffusion models throughout the remainder of this study.


\subsection{Precipitation intensity and ensemble dispersion}
\label{section:intensity-structure}

\begin{figure}[h]
    \centering
    \includegraphics[width=1.1\linewidth]{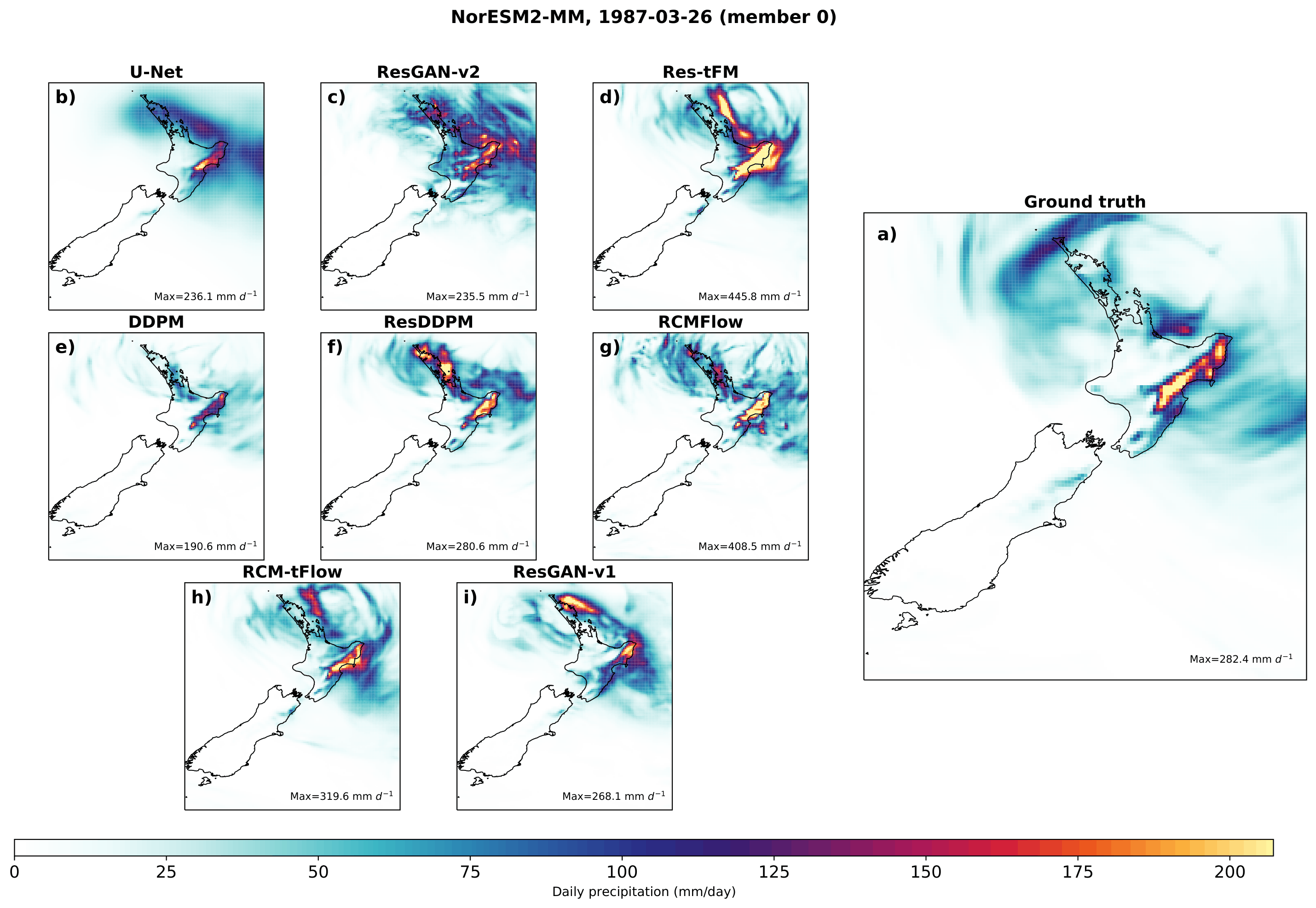}
    \caption{\footnotesize Spatial comparison of predicted daily precipitation
    fields for a simulated ex-tropical cyclone event (NorESM2-MM,
    1987-03-26).
    (a)~Ground truth CCAM field.
    (b--i)~Predictions from all eight models: U-Net, ResGAN-v2, Res-tFM,
    DDPM, ResDDPM, RCMFlow, RCM-tFlow, and ResGAN-v1.
    Maximum daily precipitation values are indicated for each panel.}
    \label{fig:casestudy}
\end{figure}

A key advantage of generative downscaling approaches over deterministic
methods is their ability to resolve fine-scale spatial structure and generate
physically plausible ensemble spread.
To illustrate this, Fig.~\ref{fig:casestudy} shows predictions from all eight
models for a simulated ex-tropical cyclone event over New Zealand
(NorESM2-MM, 1987-03-26).
The ground truth CCAM field (Fig.~\ref{fig:casestudy}a) exhibits a distinct
cyclonic precipitation structure with organised spiral banding and a maximum
of 282.4~mm~d$^{-1}$.

Compared to the ground truth (Fig.~\ref{fig:casestudy}a), the deterministic U-Net produces an overly smoothed precipitation field with no discernible cyclonic structure (Fig.~\ref{fig:casestudy}b). Both ResGAN variants resolve more fine-scale detail but similarly fail to resolve the spatial structure of the event. In comparison, all flow matching and diffusion models broadly capture the cyclonic spatial structure, though with notable intensity differences: DDPM substantially underestimates total rainfall (Fig.~\ref{fig:casestudy}e), and RCMFlow overestimates peak intensities with too fine-scale detail upon visual inspection (Fig.~\ref{fig:casestudy}g). Overall, ResDDPM, RCM-tFlow, and Res-tFM best represent the spatial structure of the event. It is important to note that while GANs have shown strong performance in representing the spatial structure of precipitation in previous case studies \citep{rampal2025reliable}, those have focused predominantly on frontal and linear precipitation systems. Their comparatively poor performance here may therefore reflect the distinct dynamics of tropical cyclones rather than a general limitation, and conclusions should be tempered given this is a single case study.

\begin{figure}[h]
    \centering
    \includegraphics[width=1.1\linewidth]{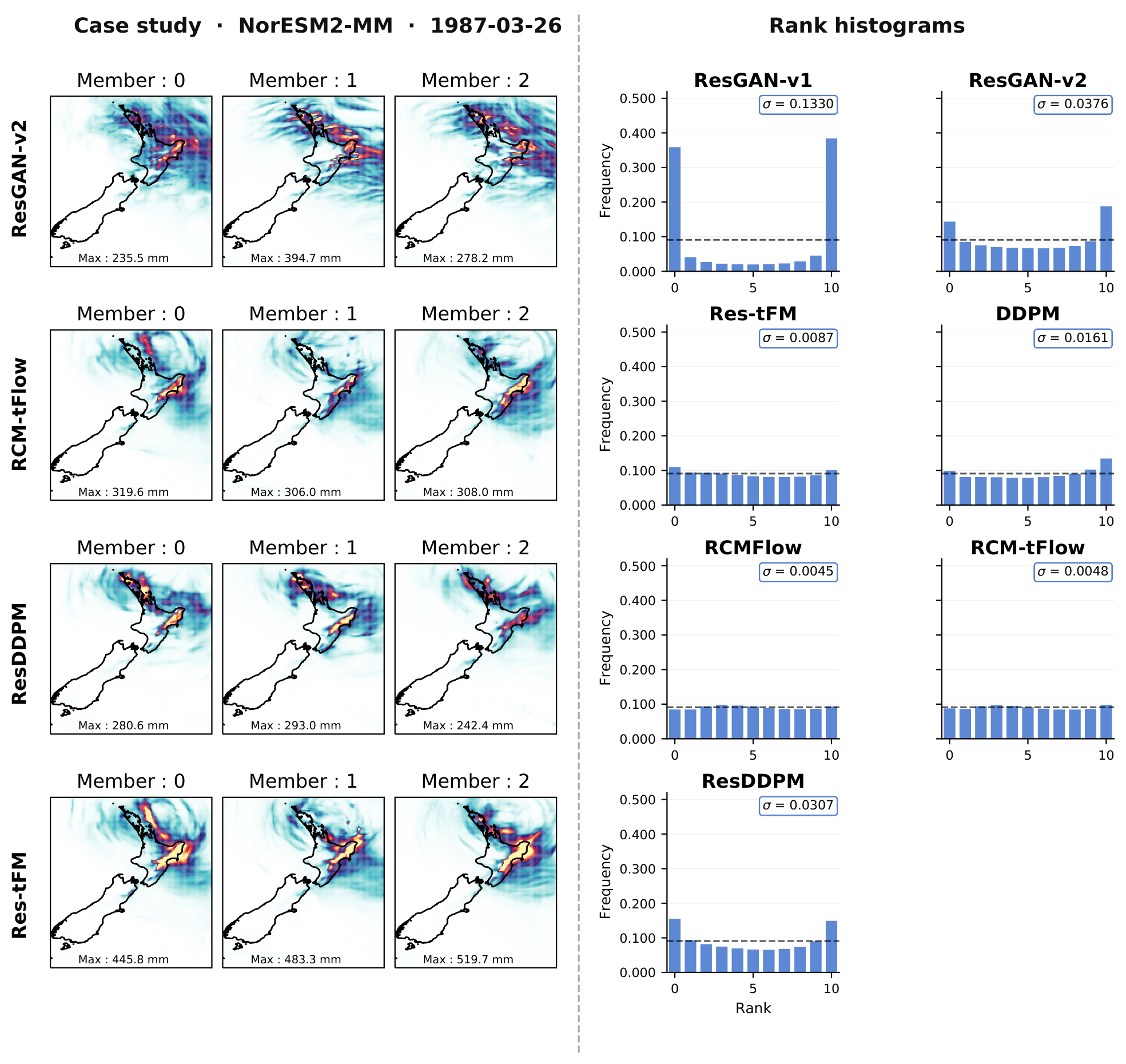}
    \caption{\footnotesize Ensemble spread for the same ex-tropical cyclone
    event (NorESM2-MM, 1987-03-26).
    Left panels show three ensemble members for ResGAN-v2, RCM-tFlow,
    ResDDPM, and Res-tFM.
    Right panels show rank histograms for all generative models, with the
    standard deviation of the rank histogram ($\sigma$) inset; a perfectly
    calibrated ensemble produces a flat rank histogram ($\sigma = 0$).}
    \label{fig:ensemble_dispersion}
\end{figure}

Figure~\ref{fig:ensemble_dispersion} illustrates ensemble spread for the same event, showing three ensemble members each for ResGAN-v2, RCM-tFlow, ResDDPM, and Res-tFM. To more quantitatively examine ensemble calibration, we constructed rank histograms from 10-member ensembles over one full year of predictions (2097) for all generative models. Rank histograms are a complementary evaluation measure to the spread-skill score (Figure~\ref{fig:sampling_speed} and Supplementary Figure S3), and are constructed by ranking the verifying ground truth RCM simulation within the predicted ensemble at each grid point and day; a perfectly calibrated ensemble yields a flat histogram ($\sigma = 0$), while U-shaped or dome-shaped departures indicate under- or over-dispersion, respectively.

We find significant differences in ensemble calibration across model families. ResGAN-v2 is slightly underdispersive ($\sigma = 0.0376$), while ResGAN-v1 is considerably more so ($\sigma = 0.1330$), with ensemble members so similar as to be nearly deterministic. In contrast, diffusion and flow matching models produce substantially better calibrated ensembles, with RCM-tFlow, RCMFlow, DDPM, and Res-tFM all exhibiting near-flat rank histograms ($\sigma \leq 0.0161$). ResDDPM is a notable exception, showing slight underdispersion relative to its diffusion and flow matching counterparts. The Continuous Ranked Probability Score (CRPS), which jointly measures ensemble spread and accuracy by evaluating the full predicted probability distribution against the observed value, provides a complementary assessment of probabilistic skill (Supplementary Figure~S3). CRPS scores are generally lowest for the flow matching models, indicating that these approaches produce the best calibrated and most skilful ensemble predictions.

\subsection{Spatial structure and climatological metrics}
\label{section:climatology-metrics}

We now assess model performance across the full historical (1985--2014) and future (2080--2099) periods for all eight configurations using metrics that evaluate spatial structure (RALSD), distributional skill (LHD), and climatological accuracy. Results for the EC-Earth GCM are presented in the figures for illustrative purposes, with averages across all four GCMs reported in Table~\ref{tab:all_metrics_multimodel}.

\begin{figure}
    \centering
    \begin{adjustbox}{width=1.15\linewidth,center}
        \includegraphics{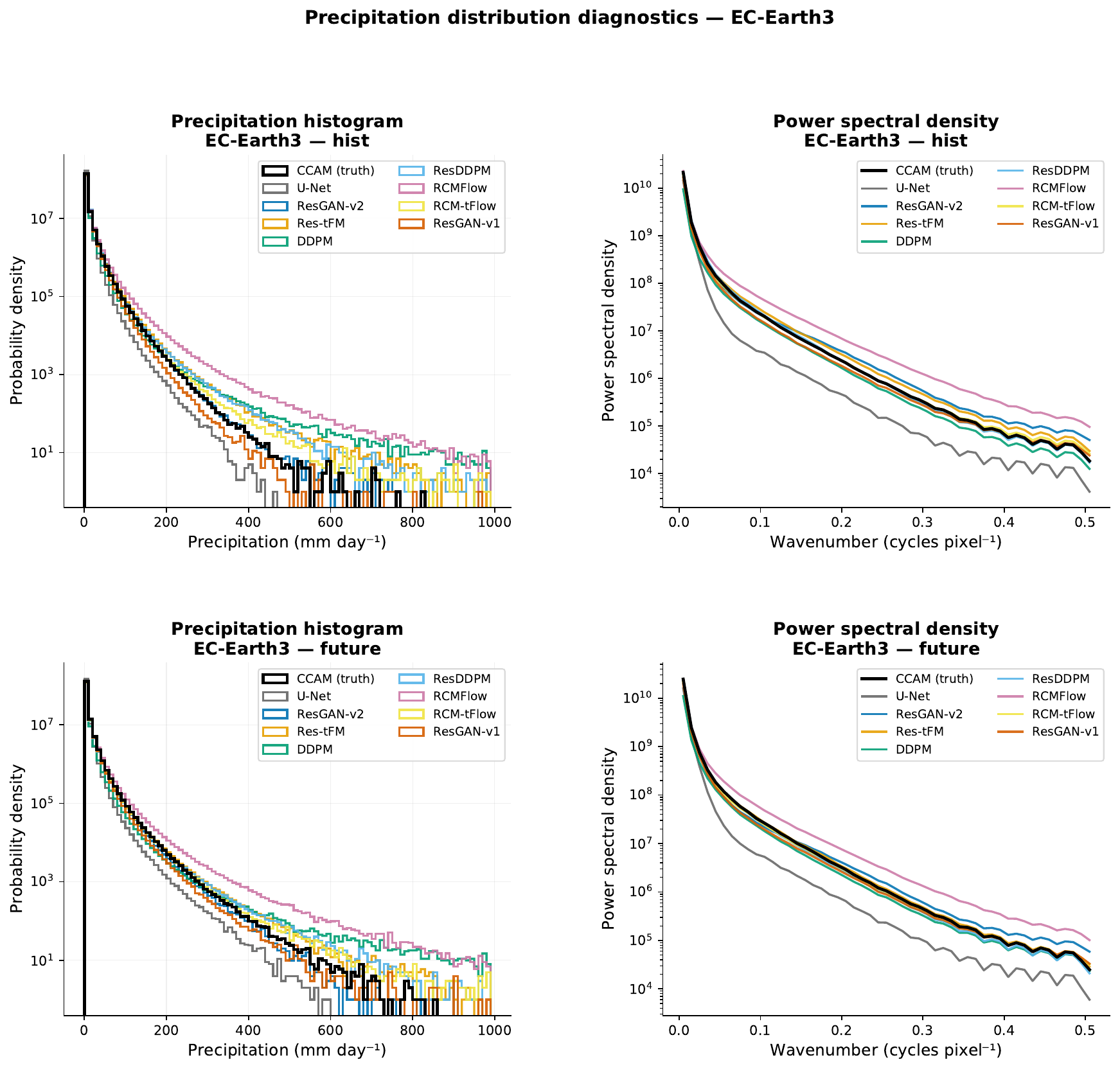}
    \end{adjustbox}
    \caption{\footnotesize Precipitation intensity distributions (left) and radially-averaged power spectral densities (right) for the historical (1985--2014, top) and future (2080--2099, bottom) periods, evaluated against EC-Earth3-driven CCAM. Intensity histograms include all days exceeding 1~mm~d$^{-1}$ and are shown on a logarithmic probability density scale. Power spectral densities are radially averaged as a function of wavenumber (cycles pixel$^{-1}$). RALSD and LHD metrics summarising these diagnostics are reported in Table~\ref{tab:all_metrics_multimodel}.}
    \label{fig:metric_LHD_RALSD}
\end{figure}

Figure~\ref{fig:metric_LHD_RALSD} (left panel) shows precipitation intensity histograms and radially-averaged power spectral densities for both historical and future periods. Most approaches reproduce the precipitation intensity distribution well up to approximately 100~mm~d$^{-1}$, beyond which performance differs significantly. In particular, the U-Net systematically underestimates precipitation frequency above ${\sim}50$~mm~d$^{-1}$, whereas most flow matching and diffusion models generally overestimate this. RCMFlow substantially overestimates event frequency above 100~mm~d$^{-1}$, whereas ResDDPM, RCM-tFlow, and Res-tFM show only modest overestimation beyond ${\sim}300$~mm~d$^{-1}$. This overestimation is most pronounced in the historical period and less so in the future, where the ground truth distribution itself shifts toward higher intensities. Across both periods, ResGAN-v2 best reproduces the target intensity distribution (LHD: 0.27~dB historical, 0.39~dB future; Table~\ref{tab:all_metrics_multimodel}), followed by ResGAN-v1 and RCM-tFlow. We note that diffusion and flow matching models that overestimate the precipitation tail can, in some instances, yield LHD scores worse than the U-Net. For the power spectral densities, the U-Net captures spectral power at small wavenumbers ($\leq 0.05$; large spatial scales) but substantially underestimates power at high wavenumbers (Figure~\ref{fig:metric_LHD_RALSD}, right panel), resulting in large RALSD scores (4.90, 5.06~dB historical and future; Table~\ref{tab:all_metrics_multimodel}). RCMFlow overestimates the spectra across most scales, consistent with the excessive spatial detail noted in Section~\ref{section:intensity-structure} and higher RALSD scores (5.01~dB historical, 4.16~dB future period). ResGAN-v1, ResDDPM, and RCM-tFlow best reproduce the CCAM power spectrum across all scales and both periods, with ResDDPM (RALSD: 0.65, 0.41~dB) and RCM-tFlow (0.68, 0.62~dB) achieving the lowest RALSD scores overall.

\begin{figure}
    \centering
    \begin{adjustbox}{width=1.15\linewidth,center}
        \includegraphics{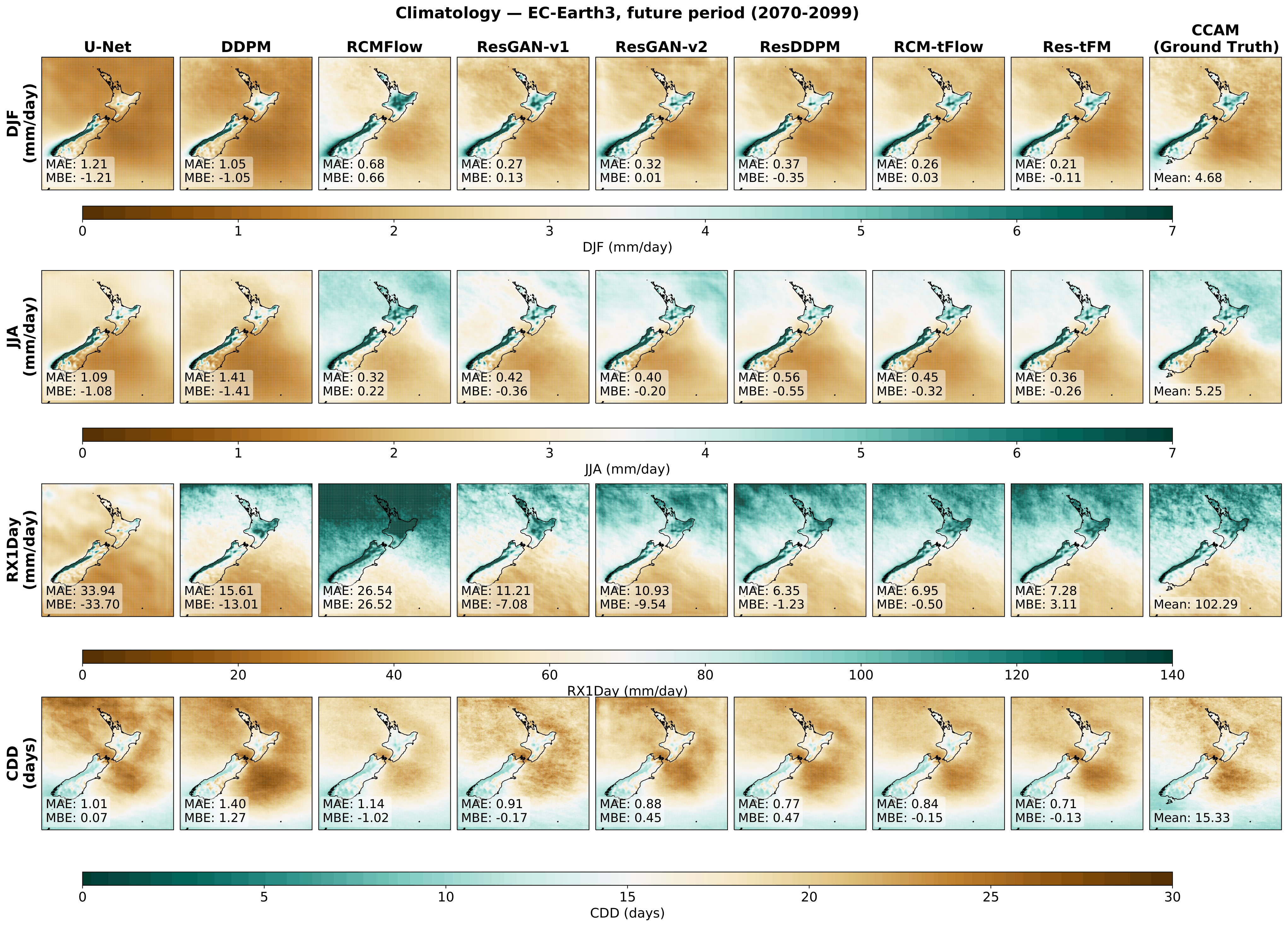}
    \end{adjustbox}
    \caption{\footnotesize
    Out-of-sample performance of all eight downscaling models in reproducing
    future-period climatologies of seasonal mean precipitation (DJF and JJA),
    annual maximum 1-day precipitation (Rx1Day), and consecutive dry days
    (CDD), evaluated against EC-Earth3-driven CCAM (2070--2099).
    Each row shows l metric; columns correspond to
    U-Net, DDPM, RCMFlow, ResGAN-v1, ResGAN-v2, ResDDPM, RCM-tFlow, and
    Res-tFM, with the CCAM ground truth in the final column.
    Mean absolute error (MAE; mm~d$^{-1}$ or days) and mean bias error (MBE)
    relative to CCAM are shown for each model; the CCAM domain-averaged mean
    is reported in the ground truth column.}
    \label{fig:metrics_all}
\end{figure}

Spatial climatologies of DJF and JJA mean precipitation, Rx1Day, and CDD for the future period under EC-Earth are presented in Figure~\ref{fig:metrics_all}. For seasonal mean precipitation, the U-Net and DDPM exhibit the largest errors and dry biases across both periods (Table~\ref{tab:all_metrics_multimodel}). Most generative configurations perform considerably better: Res-tFM and RCM-tFlow achieve consistently low errors across both seasons (e.g., future DJF: 0.31, 0.32~mm~d$^{-1}$; JJA: 0.32, 0.35~mm~d$^{-1}$, respectively), and ResGAN-v1 performs best for DJF (future MAE: 0.29~mm~d$^{-1}$) though less so for JJA. Notably, despite its poor representation of the precipitation intensity distribution, RCMFlow achieves lower seasonal mean errors than both the U-Net and DDPM. For Rx1Day, the U-Net and RCMFlow are the worst-performing configurations, with large negative and positive biases, respectively. ResDDPM, RCM-tFlow, and Res-tFM best capture the Rx1Day climatology (future MAE: 6.52, 6.35, 8.14~mm~d$^{-1}$, respectively), with the GAN variants having slightly higher MAEs. For CDD, most configurations perform reasonably well. DDPM exhibits the largest errors, while RCMFlow tends to underestimate mean CDD. ResDDPM and RCM-tFlow achieve the best performance across all configurations (future MAE: 0.93, 1.04~days, respectively).

\begin{table}[htbp]
  \centering
  \caption{Out-of-sample performance averaged across four evaluation GCMs
    (EC-Earth3, AWI-CM-1-1-MR, CNRM-CM6-1, NorESM2-MM).
    Climatological MAE is reported in mm\,d$^{-1}$ (DJF, JJA, Rx1Day)
    and days (CDD); LHD and RALSD are reported in dB.
    Bold indicates best average performance per metric and period.}
  \label{tab:all_metrics_multimodel}
  \resizebox{\textwidth}{!}{%
  \begin{tabular}{@{}ll c cc cc ccc@{}}
    \toprule
    & & \multicolumn{1}{c}{} & \multicolumn{2}{c}{GAN} & \multicolumn{2}{c}{DDPM} & \multicolumn{3}{c}{Flow Matching} \\
    \cmidrule(lr){4-5} \cmidrule(lr){6-7} \cmidrule(l){8-10}
    Metric & Period
      & U-Net & ResGAN-v1 & ResGAN-v2 & DDPM & ResDDPM & Res-tFM & RCMFlow & RCM-tFlow \\
    \midrule
    \multirow{2}{*}{CDD (days)}
      & Hist.   & 1.28 & 1.01         & 1.21 & 1.34 & 1.06         & 1.16 & 1.44 & \textbf{0.97} \\
      & Future  & 1.21 & 1.17         & 1.04 & 1.20 & \textbf{0.93} & 1.06 & 1.59 & 1.04 \\
    \addlinespace
    \multirow{2}{*}{DJF (mm\,d$^{-1}$)}
      & Hist.   & 0.87 & \textbf{0.26} & 0.47 & 1.07 & 0.40         & 0.38 & 0.60 & 0.37 \\
      & Future  & 0.96 & \textbf{0.29} & 0.42 & 0.93 & 0.35         & 0.31 & 0.69 & 0.32 \\
    \addlinespace
    \multirow{2}{*}{JJA (mm\,d$^{-1}$)}
      & Hist.   & 0.91 & 0.51         & 0.52 & 1.45 & 0.64         & 0.43 & 0.48 & \textbf{0.42} \\
      & Future  & 0.90 & 0.55         & 0.44 & 1.41 & 0.55         & \textbf{0.32} & 0.36 & 0.35 \\
    \addlinespace
    \multirow{2}{*}{Rx1Day (mm\,d$^{-1}$)}
      & Hist.   & 21.88 & 8.39        & 8.00 & 11.84 & 6.29        & 8.25  & 21.83 & \textbf{5.91} \\
      & Future  & 25.54 & 9.94        & 9.31 & 12.88 & 6.52        & 8.14  & 21.48 & \textbf{6.35} \\
    \midrule
    \multirow{2}{*}{LHD (dB)}
      & Hist.   & 1.11 & 0.64         & \textbf{0.27} & 1.51 & 1.48 & 1.36 & 2.54 & 0.74 \\
      & Future  & 1.37 & 0.65         & \textbf{0.39} & 1.21 & 1.14 & 0.94 & 2.09 & 0.71 \\
    \addlinespace
    \multirow{2}{*}{RALSD (dB)}
      & Hist.   & 4.90 & 0.72         & 2.42 & 3.52 & \textbf{0.65} & 1.59 & 5.01 & 0.68 \\
      & Future  & 5.06 & 0.73         & 1.90 & 1.36 & \textbf{0.41} & 0.85 & 4.16 & 0.62 \\
    \bottomrule
  \end{tabular}}
\end{table}


\subsection{Climate change signals}
\label{section:climate-change}

\begin{figure}[H]
    \centering
    \includegraphics[width=0.95\linewidth]{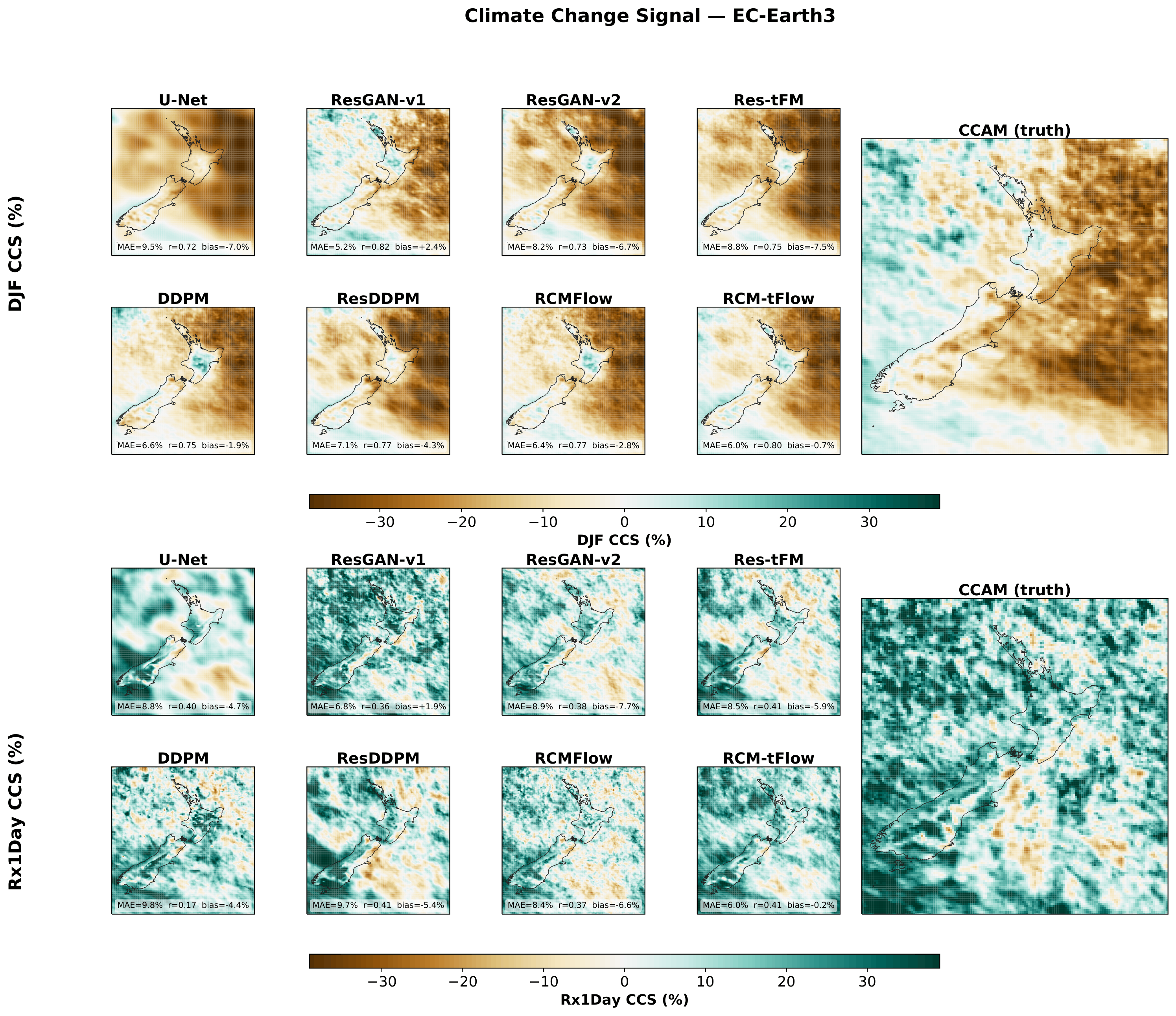}
    \caption{\footnotesize Spatial climate change signals (CCS) in seasonal
    mean precipitation (DJF; upper panel) and annual maximum daily
    precipitation (Rx1Day; lower panel) for EC-Earth3, expressed as
    percentage change between the historical (1985--2014) and future
    (2080--2099) periods.
    Predictions from all eight downscaling models are shown alongside the
    CCAM ground truth (top right of each panel).
    Mean absolute error (MAE; \%), spatial correlation coefficient ($r$),
    and mean bias (\%) relative to CCAM are shown inset for each model;
    metrics are computed over 10$\times$10-pixel pooled blocks to reduce
    sensitivity to small-scale positional errors.}
    \label{fig:cc_signal}
\end{figure}

We now assess whether models correctly reproduce the climate change signal (CCS): the relative change in precipitation between the historical (1985--2014) and future (2080--2099) periods. We focus on DJF mean precipitation and Rx1Day, as these span mean and extreme precipitation changes and represent particularly challenging targets given their inherent uncertainty \citep{lewis2025storylines, gibson_storylines_2024}. Since we are primarily interested in large-scale spatial patterns and the signals are inherently noisy, all metrics are computed over 10$\times$10-pixel spatially averaged blocks to avoid double-penalty effects. Spatial results for EC-Earth3 are shown in Figure~\ref{fig:cc_signal}, with performance averaged across all four evaluation GCMs reported in Table~\ref{tab:combined_metrics_extrmes_signal}.

For the DJF CCS (Fig.~\ref{fig:cc_signal}, upper panel), CCAM projects a predominantly drying pattern over the Eastern Ocean surrounding New Zealand, with a weak or negligible signal over much of the North Island and slight drying over the eastern portions of both islands. Most models capture the broad spatial pattern of this signal reasonably well, with spatial correlations ranging from $r = 0.72$ (U-Net) to $r = 0.82$ (ResGAN-v1). However, nearly all configurations exhibit a systematic dry bias. Averaged across all four evaluation GCMs, ResGAN-v1 achieves the lowest DJF CCS MAE (4.5\%; bias: $+2.0$\%), followed by RCM-tFlow (MAE: 5.2\%; bias: $-1.8$\%), while all other configurations show substantially stronger dry biases, including models that performed well on mean climatological metrics.

For Rx1Day and in the case of EC-Earth3 (Fig.~\ref{fig:cc_signal}, lower panel), CCAM projects a widespread increase in extreme precipitation across most of the domain, particularly over land, though spatial patterns are noisy and the signal is weak in some locations, including the eastern parts of both islands and the south-east of New Zealand. The multi-model mean across four GCMs shows that this spatial pattern is broadly consistent across all GCMs (Supplementary Figure S4). Nearly all models underestimate this signal with predominantly negative biases, which is consistent across all GCMs. RCM-tFlow and ResGAN-v1 again perform best, achieving the lowest multi-GCM MAEs (6.2\% and 6.6\%, respectively) and near-zero mean biases ($-1.7$\% and $0.0$\%), while all other configurations exhibit substantial negative biases.

\begin{table}[htbp]
  \centering
  \caption{Spatial climate change signal MAE (\%) and mean bias (\%) for
    DJF and Rx1Day, computed over 10$\times$10-pixel pooled blocks.
    \textbf{Bold} indicates best per row.
    \textit{ACCESS-CM2} is the training GCM shown for reference.}
  \label{tab:combined_metrics_extrmes_signal}
  \resizebox{\textwidth}{!}{%
  \begin{tabular}{@{}lll cccccccc@{}}
    \toprule
    & & & \multicolumn{8}{c}{Downscaling Model} \\
    \cmidrule(l){4-11}
    Metric & Stat & GCM
      & U-Net & ResGAN-v1 & ResGAN-v2 & Res-tFM & DDPM & ResDDPM & RCMFlow & RCM-tFlow \\
    \midrule
    \multirow{12}{*}{DJF CCS (\%)}
      & \multirow{6}{*}{MAE}
        & EC-Earth3    &  9.5 & \textbf{5.2} &  8.2 &  8.8 & 6.6 & 7.1 & 6.4 & 6.0 \\
      & & AWI-CM       & 10.4 & \textbf{4.0} &  9.1 &  8.6 & 6.5 & 7.3 & 6.4 & 5.2 \\
      & & CNRM-CM6     & 13.7 & \textbf{4.7} & 11.7 & 10.8 & 5.6 & 7.2 & 6.1 & 5.3 \\
      & & NorESM2      &  5.9 & \textbf{4.3} &  5.5 &  5.2 & 5.1 & 4.5 & 4.5 & 4.3 \\
    \addlinespace[1pt]
      & & \textbf{Avg} &  9.9 & \textbf{4.5} &  8.6 &  8.4 & 5.9 & 6.5 & 5.8 & 5.2 \\
      & & \textit{ACCESS-CM2} & 16.8 & \textbf{3.5} & 12.9 & 14.0 & 9.4 & 10.2 & 8.4 & 6.5 \\
    \addlinespace[4pt]
      & \multirow{6}{*}{Bias}
        & EC-Earth3    &  $-7.0$ & $+2.4$ & $-6.7$ & $-7.5$ & $-1.9$ & $-4.3$ & $-2.8$ & \textbf{$-0.7$} \\
      & & AWI-CM       &  $-9.8$ & \textbf{$+0.6$} & $-8.5$ & $-7.8$ & $-4.5$ & $-6.3$ & $-5.2$ & $-3.4$ \\
      & & CNRM-CM6     & $-14.0$ & $+2.7$ & $-11.8$ & $-10.9$ & \textbf{$-1.5$} & $-6.5$ & $-5.2$ & $-2.2$ \\
      & & NorESM2      &  $-5.6$ & $+2.4$ & $-4.6$ & $-3.9$ & $-1.0$ & $-3.1$ & $-2.2$ & \textbf{$-0.6$} \\
    \addlinespace[1pt]
      & & \textbf{Avg} &  $-9.1$ & $+2.0$ & $-7.9$ & $-7.5$ & $-2.2$ & $-5.0$ & $-3.8$ & \textbf{$-1.8$} \\
      & & \textit{ACCESS-CM2} & $-17.3$ & \textbf{$+0.7$} & $-13.2$ & $-14.4$ & $-7.7$ & $-10.6$ & $-8.3$ & $-5.5$ \\
    \midrule
    \multirow{12}{*}{Rx1Day CCS (\%)}
      & \multirow{6}{*}{MAE}
        & EC-Earth3    &  8.8 & 6.8         &  8.9 &  8.5 &  9.8 &  9.7 & 8.4 & \textbf{6.0} \\
      & & AWI-CM       & 10.4 & \textbf{6.6} &  9.5 & 10.4 &  9.6 &  9.9 & 9.5 & 7.3 \\
      & & CNRM-CM6     & 10.4 & 7.1         &  9.0 & 10.6 &  8.1 &  8.3 & 8.7 & \textbf{5.8} \\
      & & NorESM2      &  6.4 & \textbf{5.9} &  6.1 &  6.8 &  6.8 &  6.7 & 6.2 & 5.9 \\
    \addlinespace[1pt]
      & & \textbf{Avg} &  9.0 & 6.6         &  8.4 &  9.1 &  8.6 &  8.7 & 8.2 & \textbf{6.2} \\
      & & \textit{ACCESS-CM2} & 16.3 & \textbf{5.6} & 12.9 & 14.6 & 12.0 & 11.6 & 10.8 & 8.0 \\
    \addlinespace[4pt]
      & \multirow{6}{*}{Bias}
        & EC-Earth3    & $-4.7$ & $+1.9$ & $-7.7$ & $-5.9$ & $-4.4$ & $-5.4$ & $-6.6$ & \textbf{$-0.2$} \\
      & & AWI-CM       & $-9.1$ & \textbf{$-1.5$} & $-8.5$ & $-9.2$ & $-7.9$ & $-8.0$ & $-9.1$ & $-5.2$ \\
      & & CNRM-CM6     & $-9.5$ & \textbf{$-0.2$} & $-8.1$ & $-9.1$ & $-4.5$ & $-5.1$ & $-7.5$ & $-0.9$ \\
      & & NorESM2      & $-0.9$ & \textbf{$-0.2$} & $-3.5$ & $-3.0$ & $-4.2$ & $-2.7$ & $-4.1$ & $-0.5$ \\
    \addlinespace[1pt]
      & & \textbf{Avg} & $-6.0$ & \textbf{$+0.0$} & $-6.9$ & $-6.8$ & $-5.3$ & $-5.3$ & $-6.8$ & $-1.7$ \\
      & & \textit{ACCESS-CM2} & $-16.6$ & \textbf{$+0.9$} & $-12.7$ & $-14.2$ & $+13.1$ & $-11.4$ & $-10.1$ & $-5.1$ \\
    \bottomrule
  \end{tabular}}
\end{table}

\section{Discussion}\label{section:discussion}

An important practical consideration for the operational deployment of diffusion and flow matching models, for example, for downscaling CMIP6 or CMIP7 multi-model ensembles — is inference cost. Standard DDPM implementations require approximately 1000 sampling steps using a probabilistic Markovian sampling process \citep{ho2020denoising, addison2026machine}, imposing substantial computational overhead relative to GANs, which require only a single forward pass. We demonstrate that just 25 sampling steps, using the AB2-PC solver for flow matching and DDIM for diffusion, yield well-calibrated, skilful predictions with little compromise in performance relative to higher step counts. Notably, AB2-PC incurs no additional network evaluations relative to a first-order Euler solver, as it reuses the velocity field computed at the previous timestep; this reduces inference cost by approximately 40-fold relative to 1000-step approaches used in prior work \citep{addison2026machine, legasa_regional_2026}.

By using 25 inference steps, non-residual and residual flow matching and diffusion models require approximately 31 and 61 times the inference cost of the U-Net baseline, respectively (Table~\ref{tab:gpu_hours}), compared with 2--3 times for GAN variants. For applications requiring downscaling of large climate model ensembles \citep[e.g.,][]{rampal2025downscaling, lewis2025generative}, GANs remain highly practical, capable of generating 10,000 years of output on a single A100 GPU in approximately one day \citep{rampal2025downscaling}. At 25 steps, flow matching and diffusion models would require approximately one week, though this remains a substantial improvement over dynamical downscaling and can be reduced further through multi-GPU or parallelised inference. As for training costs, GAN variants are considerably more expensive to train (47--67~GPU hours per 200 epochs) than diffusion and flow matching models (13--24~GPU hours; Table~\ref{tab:gpu_hours}), partly because the GAN discriminator requires multiple updates per generator step. Overall, the perception that diffusion and flow matching models are prohibitively expensive relative to deterministic approaches and GANs \citep{rampal2024enhancing} should be less of a concern with the adoption of higher-order solvers and reduced step counts, as demonstrated here. Model selection should therefore be guided primarily by task-specific skill requirements rather than computational cost alone. For example, climatological accuracy and reliable climate change signal reconstruction may be the primary criteria for large ensemble downscaling applications, whereas ensemble calibration and spread may be more important for weather-scale applications.

\begin{table}[htbp]
  \centering
  \caption{\footnotesize Training and inference cost for all models evaluated in this study.
    Training time is reported in GPU hours normalised to 200 epochs.
    Inference time is reported in seconds per year (365 days) of output generated per ensemble member,
    and expressed relative to the U-Net baseline.
    Residual models (ResDDPM, Res-tFM) include the U-Net inference cost.}
  \label{tab:gpu_hours}
  \resizebox{\textwidth}{!}{
  \begin{tabular}{lcccccccc}
    \hline
    & \textbf{U-Net} & \textbf{ResGAN-v1} & \textbf{ResGAN-v2} & \textbf{DDPM} & \textbf{ResDDPM} & \textbf{RCMFlow} & \textbf{RCM-tFlow} & \textbf{Res-tFM} \\
    \hline
    Training [GPU hrs / 200 epochs] &  6  & 47  & 67  & 13  & 24  & 13  & 13  & 24  \\
    Inference [s/yr/member]         & 0.3 & 1.1 & 0.7 & 10.5 & 21.0 & 10.5 & 10.5 & 21.0 \\
    Inference (w.r.t.\ U-Net)       & 1$\times$ & 3$\times$ & 2$\times$ & 31$\times$ & 61$\times$ & 31$\times$ & 31$\times$ & 61$\times$ \\
    \hline
  \end{tabular}
  }
\end{table}

Consistent with previous studies \citep[e.g.,][]{rampal2025reliable, addison2026machine, doury_suitability_2024, legasa_regional_2026}, the deterministic U-Net produces spatially smooth predictions, systematically underestimates extremes, and performs worst across most metrics examined in this study. All generative configurations improve on this baseline across nearly all metrics, though no single model performs consistently well across both weather-related metrics (ensemble calibration and dispersion) and climate downscaling metrics (climatological accuracy and climate change signal reproduction). Flow matching models are most skilful across weather-related metrics, whereas diffusion models produce well-calibrated ensembles but with higher CRPS, and GAN variants generally produce underdispersive ensembles. Additionally, all diffusion and flow matching models generate more physically realistic spatial structure than the GAN variants, as illustrated by the ex-tropical cyclone case study. Overall, GANs appear less well suited to ensemble generation and weather applications compared to flow matching and diffusion approaches, and exhibit notable limitations in representing the dynamics of certain weather events.

Most generative configurations reproduce the ground truth RCM climatology well across all climatological metrics (DJF and JJA seasonal mean precipitation, Rx1Day, and CDD), with GAN, diffusion, and flow matching variants performing competitively across most of these metrics. A notable exception is the representation of the precipitation intensity distribution, where GAN configurations consistently outperform diffusion and flow matching models (Table~\ref{tab:all_metrics_multimodel}); most diffusion and flow matching models slightly overestimate the frequency of very rare, high-intensity precipitation events. The better performance of GANs on this metric is likely a consequence of an intensity constraint in their loss functions \citep{rampal2025reliable}, that directly penalises departures from the ground truth intensity distribution. Most models also broadly capture the spatial pattern of the DJF climate change signal, though nearly all exhibit a systematic dry bias. Also, nearly all configurations systematically underestimate the future climate change signal of extreme precipitation (Rx1Day), despite most models being trained on data spanning the future period.

Among the eight models evaluated, only ResGAN-v1 and RCM-tFlow reliably reproduce the Rx1Day and DJF climate change signals across all evaluation GCMs, despite performing comparably to other configurations on most other metrics. Attributing this skill to specific model characteristics is challenging, as both configurations share largely identical architectures and hyperparameters with their respective model families; Res-tFM, for example, differs from RCM-tFlow only in its residual rather than direct formulation, yet performs less skilfully in reproducing climate change signals. We investigated sensitivity to training epoch and found little variation (not shown), partly owing to EMA weighting in the diffusion and flow matching configurations. Several factors may contribute: for ResGAN-v1, the intensity constraint and single-member loss formulation may encourage more faithful representation of the tail response under warming; for RCM-tFlow, the Student-$t$ noise prior may better represent the heavy-tailed precipitation distribution \citep{pandey2024heavy}. These explanations are largely speculative, however, and it is possible that skill on this metric emerges partly by chance, given that performance on conventional metrics does not appear to be a reliable predictor of climate change signal skill. This result is also consistent with findings from a previous study focused on diffusion models \citep{addison2026machine} and a recent intercomparison benchmarking over 40 models across three regions \citep{rampal2026cordexmlbench}, where most models underestimated the climate change signal of extreme precipitation. This challenge is likely a result of the limited number of training samples available to learn the extreme precipitation responses, as extremes are rare by definition, and the inherently noisy nature of the signal itself. Developing training strategies that explicitly target climate change signal fidelity, for example by monitoring this as a diagnostic during training, may help identify more skilful models, and further underscores the importance of explicitly evaluating climate change signals in downscaling assessments \citep{rampal2024enhancing, kendon_potential_2025}.

Overall, ResGAN-v1 and RCM-tFlow emerge as the most suitable configurations for downscaling large ensembles of climate projections. ResGAN-v1 is best suited to workflows where producing climatologies and climate change signals are the primary objectives rather as a pose to weather generation, and has already been applied to produce downscaled datasets used to study rare extremes, drought, and internal variability \citep{lewis2025generative, rampal2025downscaling}. In comparison, RCM-tFlow offers a is more well rounded in terms of performance,e, making it the preferred choice where both weather- and climate-scale skill are required. Should these models perform comparably in the imperfect framework, as suggested for ResGAN-v1 by \citet{rampal2025downscaling}, both approaches are well positioned for downscaling large ensembles of climate projections. Model selection should therefore be guided by the intended application.

Several limitations of this study merit discussion. First, the intensity and statistical constraints incorporated into the GAN variants \citep{rampal2025reliable} could not be straightforwardly extended to diffusion and flow matching models; attempts to do so did not yield skilful results, likely because the noise-prediction training objective complicates the incorporation of such constraints without introducing artefacts \citep{mardani2025residual}. Instead, we focused on residual and non-residual configurations and different noise prior distributions. Exploring noise-level-modulated loss functions and the incorporation of physical constraints into diffusion and flow matching frameworks are important avenues for future research. This study also employed the perfect model framework, in which coarsened RCM fields are used as predictors rather than GCM inputs directly, in order to isolate extrapolation skill from the additional complexity of domain transfer between GCM and RCM input spaces. Evaluating these approaches in the imperfect framework, where models are applied directly to GCM inputs \citep{rampal2026cordexmlbench}, and extending the intercomparison to other variables and regions are important directions for future work.

\section{Conclusion}\label{section:conclusion}
Here we have presented a comprehensive intercomparison of generative downscaling algorithms—two GAN variants, two diffusion models, and three flow matching configurations—to assess their potential suitability for downscaling large climate model ensembles (e.g., CMIP6 or CMIP7), benchmarked against a deterministic deep learning baseline. Consistent with previous studies, we found that the deterministic model regresses to the mean, underestimates extremes and fine-scale detail, and performs poorly across most metrics. A key practical limitation of diffusion and flow matching models relative to GANs is their inference cost, as they require many sampling steps to generate predictions. We demonstrate that higher-order solvers, such as the Adams-Bashforth 2 predictor-corrector, can achieve well-calibrated and skilful predictions in as few as 25 steps. By leveraging memory from previous timesteps, this approach incurs no additional compute over first-order solvers, substantially reducing the computational overhead that has historically made diffusion-based methods expensive relative to GANs \citep{rampal2024enhancing}.

Our findings demonstrate that each generative family has distinct strengths and weaknesses, and that no single model performs well across all metrics. Diffusion and flow matching models produce more physically realistic spatial structure and better-calibrated ensembles than GANs, with flow matching achieving the best ensemble calibration and accuracy overall, making them well suited to weather-related downscaling. In contrast, GANs are underdispersive and exhibit higher CRPS, but better capture the overall precipitation distribution. Importantly, while most generative models perform well on climatological and weather metrics, all but two underestimate the climate change response of precipitation extremes, demonstrating that skill in conventional metrics and producing visually realistic predictions does not necessarily translate into reliable performance across applications, a finding that has also been echoed in a recent study \citep{rampal2026cordexmlbench}.

For applications requiring large ensembles of downscaled climate projections, ResGAN-v1 and RCM-tFlow (flow matching with a heavy-tailed prior) emerge as the most skillful approaches, owing to their strong performance in both present-day and future climates and their ability to reliably reproduce the climate change response of extremes, which is particularly relevant for studies investigating the internal variability of rare extremes \citep[e.g.,][]{rampal2025downscaling}. ResGAN-v1 remains highly computationally efficient and well suited where climatological and climate change signal measures are important, whereas RCM-tFlow offers a much more well-rounded alternative where ensemble calibration and physical realism are also required, at modestly higher inference cost. 

Overall, our study shows that reliably reproducing the climate change response of extremes is inherently challenging: models are trained to predict downscaled output on a daily basis, whereas the climate change signal emerges over decades, requiring a model to capture both daily variability and decadal trends simultaneously, and for extremes, this signal contains only a few samples from which to learn. Notably, both ResGAN-v1 and RCM-tFlow were developed in part with this objective in mind, yet why these two configurations outperform the others remains unclear. Configurations such as ResGAN-v1 have repeatedly captured climate change signals well across several studies \citep{rampal2024extrapolation, rampal2025downscaling}, likely owing to the intensity constraint used to improve extrapolation performance \citep{rampal2025reliable}, whereas RCM-tFlow's skill may stem in part from its heavy-tailed noise prior. The diffusion and flow matching models otherwise lacked such a constraint, as it could not be readily incorporated, which may have limited their ability to reproduce the climate change response of extremes. If physical constraints can be incorporated into these frameworks — for example, through noise-level-modulated loss functions or by monitoring the climate change signal during training — these models may equal or surpass GANs across all metrics while retaining their advantages in spatial structure and ensemble calibration. Future work is needed to develop these approaches and to understand why only certain configurations reliably produce climate change responses that are robust and repeatable across regions and benchmarks \citep{rampal2026cordexmlbench}, ensuring their suitability for climate risk assessment and adaptation planning. More broadly, our study underscores that standard evaluation metrics may not identify models that are skilful for climate change signals, highlighting the need for benchmarking across a diverse set of metrics like those applied here.

\section*{Open Research Section}
The RCM emulator code and datasets supporting this study are available on \href{https://zenodo.org/records/13755688}{Zenodo}. The code for training the RCM emulator is available on \href{https://github.com/nram812/An_intercomparison_of_generative_machine_learning_methods_for_downscaling_precipitation}{GitHub}.

\acknowledgments
N.R, P.B.G, Y-S.K, H-Y.L received funding from the New Zealand MBIE Endeavour Smart Ideas Fund
(C01X2202, NIW2504). N.R, T.M and P.B.G would also like to acknowledge support from MBIE
Strategic Science Investment Fund (SSIF). S.C.S would like to acknowledge the support of the Australian Research Council 21st Cen-
tury Weather (CE230100012). The authors would also like to acknowledge the New Zealand
eScience Infrastructure for providing access to GPUs.

\textnormal{\bibliography{references}}

\includepdf[pages=-]{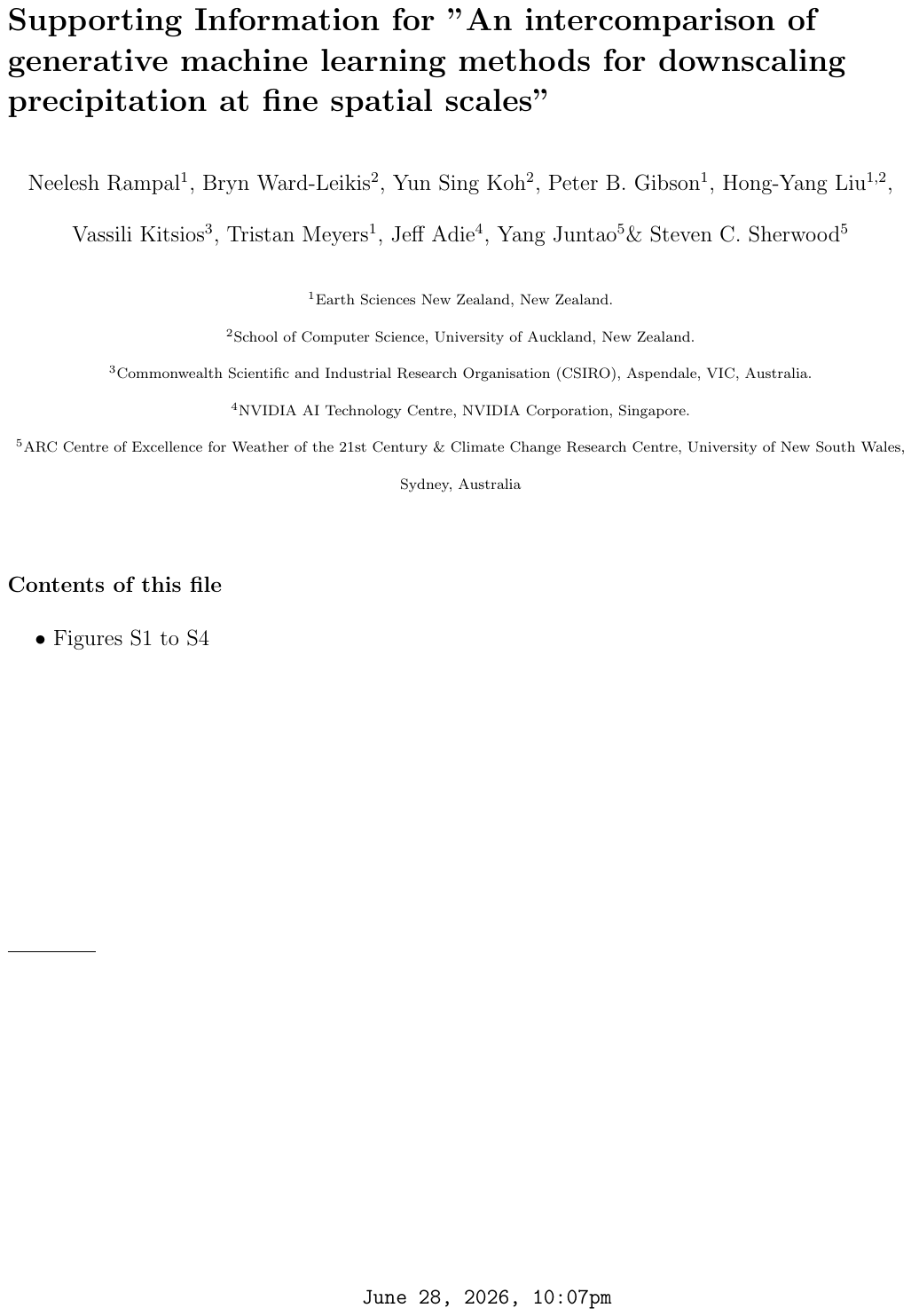}

\end{document}